\documentclass[5p]{elsarticle}
\usepackage[latin9]{inputenc}
\usepackage{color}
\usepackage{float}
\usepackage{mathrsfs}
\usepackage{amsmath}
\usepackage{amssymb}
\usepackage{graphicx}

\usepackage{stmaryrd}

\makeatletter

\usepackage{mathrsfs}

\usepackage{moreverb}
\usepackage{array}\usepackage{epsfig}\usepackage{amsthm}\usepackage{mathrsfs}

\theoremstyle{plain}

\theoremstyle{definition}

\theoremstyle{definition}

\theoremstyle{plain}

\theoremstyle{remark}

\theoremstyle{remark}

\numberwithin{equation}{section}

\newcommand{\BibTeX}{{\rmfamily B\kern-.05em \textsc{i\kern-.025em b}\kern-.08em
T\kern-.1667em\lower.7ex\hbox{E}\kern-.125emX}}

\makeatother

\begin{document}

\begin{frontmatter} 

  \title{Controllability of the Voter Model: an information theoretic approach}

\author{Pierre-Alain Toupance$^{b}$,  Laurent Lef{\`e}vre$^{b}$, and Bastien
  Chopard$^{a}$}

\address{$^{a}$ University of Geneva,  Geneva, Switzerland}

\address{$^{b}$ Univ. Grenoble Alpes, LCIS, F-26902, Valence, France}

\begin{abstract}
  We address the link between the controllability or observability of
  a stochastic complex system and concepts of information
  theory. We show that the most influential degrees of freedom can be
  detected without acting on the system, by measuring the {\em time-delayed
  multi-information}. Numerical and analytical results support this
  claim, which is developed in the case of a simple stochastic model
  on a graph, the so-called voter model. The importance of the noise
  when controlling the system is demonstrated, leading to the concept of
  {\em control length}. The link with classical control theory is given, as
  well as the interpretation of controllability in terms of the
  capacity of a communication canal.
\end{abstract}

\begin{keyword}
reachability and observability analysis \sep Mutual and multi-information \sep voter model
\end{keyword}

\end{frontmatter}

\section{Introduction}

Causality is an important concept in many areas of
science~\cite{pearl09}.  It helps to better understand the behavior of
complex dynamical systems. In particular, it reveals how the different
degrees of freedom of a system influence each other. In this paper we
investigate how causality (considered in a pragmatic and intuitive
way) can be used to discover efficient control strategies in a complex system.

The key ingredient of our approach is the concept of the most
influential components in a complex system. This notion is defined
here as the impact of controlling a given variable on the behaviour of
the other variables. For instance, one can measure the change in the
joint probability distribution (Kulback-Leibler divergence) when the
value of a selected variable is imposed. Alternatively, we can measure the
variation of an average quantity when a perturbation is
applied. The variable for which this change is the most important is
labelled as the most influential. Following this procedure we can rank
the degrees of freedom of a system from the most to the less
influential. Arguably the notion of influence depends on the quantity
used to measure the effect of forcing the variable. Then, to control
this quantity in the system, it will be more effective to act on the
corresponding most influential nodes.

The aforementioned procedure is intrusive in the sense that it
requires to act on the system to be able to determine the effects of a
perturbation. Here we would like to consider a non-intrusive approach,
essentially based on the observation of the system.  The non-intrusive
approach we proposed is based on a time delayed multi-information
measure on the free system. This procedure can be performed by simple
sampling on the system variables, even if the underlying dynamics is
unknown, like for instance in financial systems.

For now we consider, as a benchmark system, the so-called voter model
described in the next section. The most influential nodes can be
determined by controlling successively each variables and measuring
the impact on the average opinion of the entire group. We will show
that the same ranking of influence can also be obtained by monitoring
the time-delayed multi-infomation.

The determination of the most influential variables has a clear
connection with the well developed theory of control, in which
observability and controllability of a system are defined and
explored. In section~\ref{sect:controle} we make the link between the
standard concepts of control theory and our present approach. A
important element of our discussion is related to the effect of noise
on the possibility to control a system. The voter model shows that in
presence of noise the influential nodes cannot force the opinion of
the far enough agents, despite the existence of a connecting
path. This result shows the limit of some previous approaches about
the controllability of systems on a complex network~\cite{Liu11}.

The paper is organized as follows: section~\ref{sect:voter} introduces
our voter model, then section~\ref{sect:influence} demonstrates the
link between influence and time-delayed
multi-information. Section~\ref{sect:1D} solves the 1D voter model
analytically, in the meanfield regime and gives a formal link between
influence and delayed multi-information. The link between the control
length and the capacity of a communication channel is also
given. Section~\ref{sect:controle} proposes a formulation of the 1D
voter model in the usual framework of control theory. Loss of
controllability is related to the noise intensity and the cost of
controllability is expressed with a Gramian.

\section{Voter Model}\label{sect:voter}

Simple models that abstracts the process of opinion formation have
been proposed by many
researchers~\cite{RevModPhys.81.591,BC-galam:98}.  The version we
consider here is an agent-based model defined on a graph of arbitrary
topology, whether directed or not.

A binary agent occupies each node of the network. The dynamics is
specified by assuming that each agent $i$ looks at every other agent
in its neighborhood, and counts the percentage $\rho_i$ of those which
are in the state $+1$ (in case an agent is linked to itself, it
obviously belongs to its own neighborhood). A function $f$ is
specified such that $0\le f(\rho_i)\le 1$ gives the probability for
agent $i$ to be in state $+1$ at the next iteration. For instance, if
$f$ would be chosen as $f(\rho) = \rho$, an agent for which all neighbors
are in state $+1$ will turn into state $+1$ with certainty.  The
update is performed synchronously over all $n$ agents.

Formally, the dynamics of the voter model can be express as
\begin{equation}
s_i(t+1)=\left\{ \begin{array}{cc}
                  1 & \mbox{with probability $f(\rho_i(t))$} \\
                  0 & \mbox{with probability $1-f(\rho_i(t))$}\\ 
                 \end{array}
\right.
\label{eq:votermodel}
\end{equation}
where $s_i(t)\in\{0,1\}$ is the state of agent $i$ at iteration $t$,
and 
\begin{equation}
\rho_i(t)={1\over |N_i|}\sum_{j\in N_i} s_j(t).
\end{equation}
The set $N_i$ is the set of agents $j$ that are neighbors of agent $i$,
as specified by the network topology.

The global density of all $n$ agents with opinion $1$ is obviously
obtained as
\begin{equation}
  \rho(t)={1\over n}\sum_{i=1}^n s_i(t)
  \label{eq:rho}
\end{equation}

\begin{figure}[htb]
  \begin{center}
    \includegraphics[width=.3\textwidth]{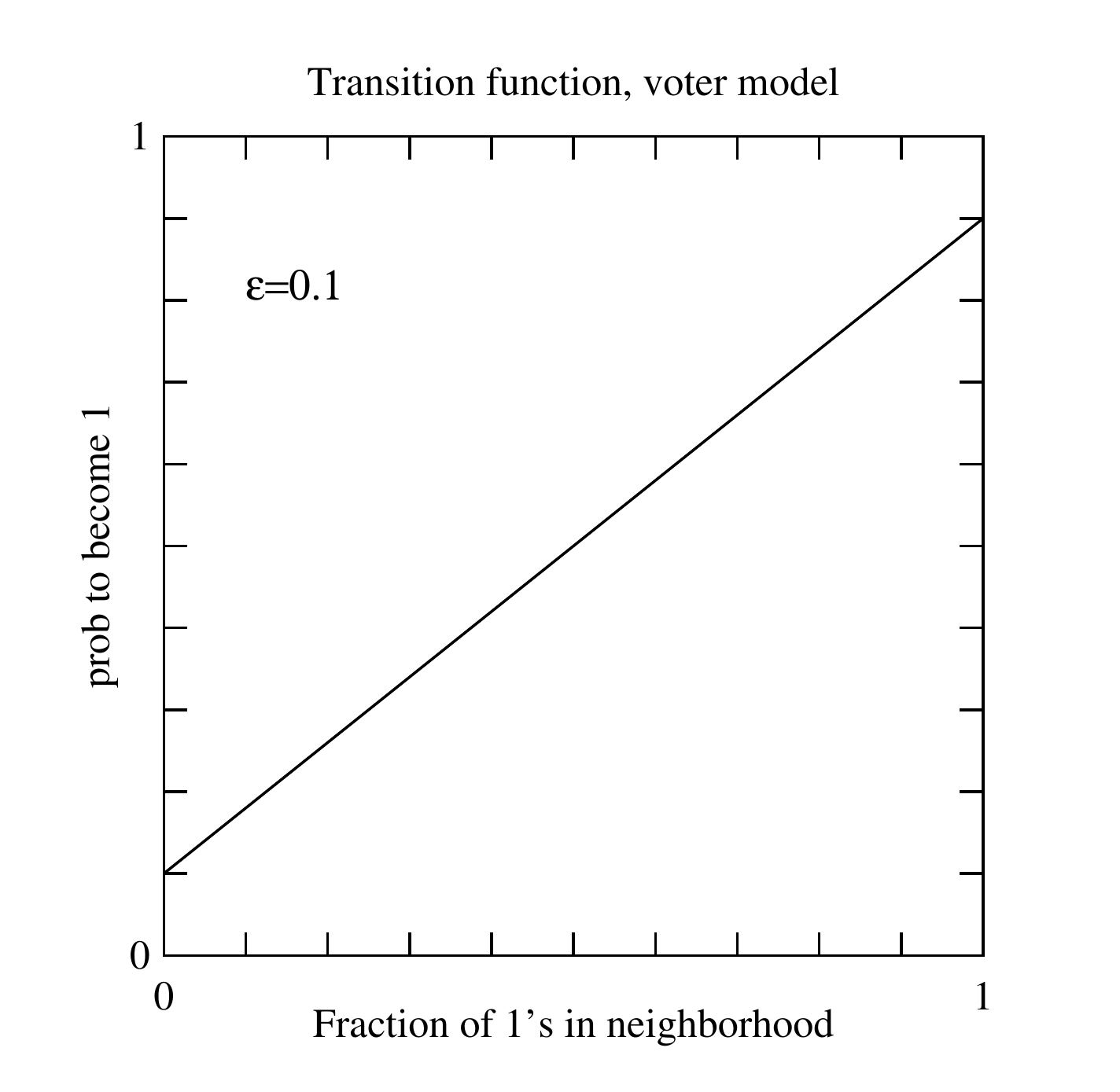}
  \end{center}
  \caption{The probablility $f(\rho)$ used in this study. The noise $\epsilon$
    is visible as the values $f(0)$ and $1-f(1)$.}
  \label{fig:plot-f}
\end{figure}

In what follows, we will use a particular function $f$, (see
Fig.~\ref{fig:plot-f})
\begin{equation}
 f(\rho)= (1- \epsilon) \rho + \epsilon (1-\rho)=(1-2\epsilon)\rho + \epsilon 
\end{equation}
The quantity $0\le\epsilon\le1/2$ is called the noise. It reflects the
probability to take a decision different from that of the
neighborhood.

To illustrate the behavior of this model, we consider a random
scale-free graph $G$, as simple instance of a social
network~\cite{Barabasi2000}. We use the algorithm  of Béla Bollobás (\cite{Bollobas2003}) to generate this graph.
 
\begin{figure}
\begin{center}
\includegraphics[width=.5\textwidth]{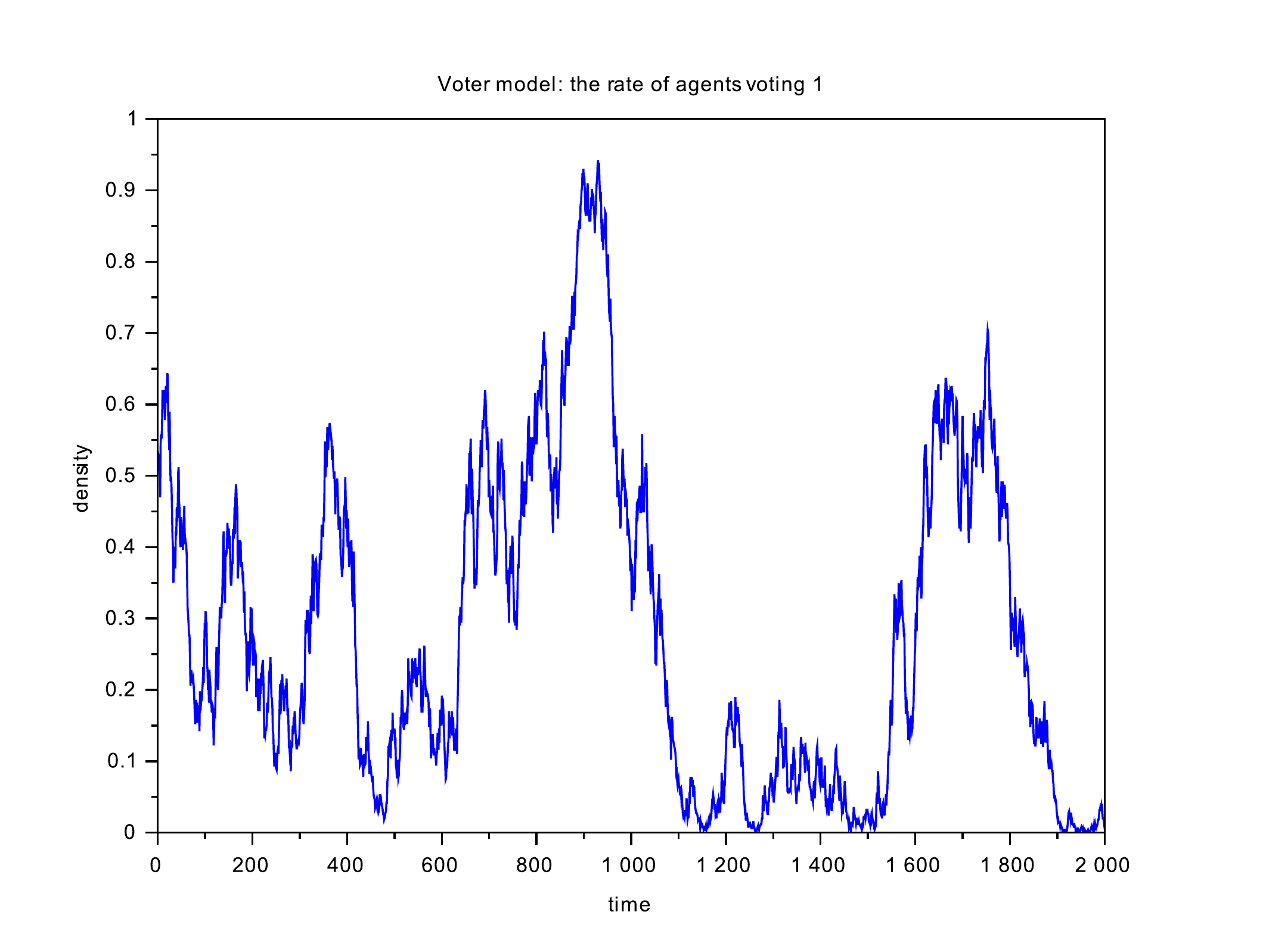}
\end{center}
\caption{Graph of the time evolution of the density of opinion 1 with
  noise $\epsilon=0,001$ and $n=200$ agents connected through a
  scale-free network.}
\label{fig:simu1}
\end{figure}

Figure~\ref{fig:simu1} shows the corresponding density of agents with
opinion 1, as a function of time. We can see that there is a lot of
fluctuations due to the fact that states ``all 0's'' or ``all 1's'' are no
longer absorbing states when $\epsilon\ne0$.

\section{Characterisation of the influence of an agent}\label{sect:influence}

In this study we would like to characterize how the opinion of one
agent influences that of its neighbors and that of the entire system.
We will first propose an approach based on information theory, and
then measure the influence directly by forcing (or controlling) the
opinion of one agent. We will show that both characterizations are
strongly correlated. The information theoretic quantities that will be
considered are the time-delayed mutual information and the
time-delayed multi-information. The purpose of considering a time
delay is to capture the causal effect of one element on another.

\subsection{Delayed mutual- and multi-information}

Let us consider a set of random variables $X_i(t)$ associated with
each agent $i$, taking their values in a set $A$. For instance,
$X_i(t)=s_i(t)$ would be the opinion of agent $i$ at iteration $t$.

To measure the influence between agents $i$ and $j$, we define the
$\tau$-delayed mutual information $w_{i,j}$ as
\begin{eqnarray}
w_{i,j}(t,\tau)&=&I(X_i(t),X_j(t+\tau))  \\
&=&
\sum_{(x,y) \in A^2} p_{xy} \log \Big( \dfrac{p_{xy}}{p_x p_y} \Big)
\label{info_mutual}
\end{eqnarray}
with 
\[ p_{xy}= \mathbb{P} (X_i(t) =x,X_j(t+\tau)=y) \]
\[ p_x = \mathbb{P} (X_i(t) =x) \text{ and } p_y= \mathbb{P} (X_j(t+\tau)=y)\]

We also define the $\tau$-delayed multi-information $w_{i}$ to
measure the influence of one agent $i$ on all the others
\begin{eqnarray}
w_{i}(t,\tau)&=&I(X_i(t),Y_i(t+\tau))
\label{info_multi}
\end{eqnarray}
\begin{equation}
Y_i(t+\tau) = \sum_{ k \neq i} X_k(t+\tau)
\end{equation}

These information metrics can be computed by the method of
sampling. We consider $N=10^5$ instances of the system in order to
perform an ensemble average.
According to the central limit theorem, we know that, with this number of instances, we obtain a precision of $3 \times 10^{-2}$ with a risk of $5 \%$ for the approximate values of the probabilities that we compute (see  \ref{appendix:Central_limit} for details).

\subsection{Non-intrusive characterisation of the nodes influences: delayed multi-information}\label{sect:non-intrusive}



\begin{figure}
\begin{center}
 \psfig{figure=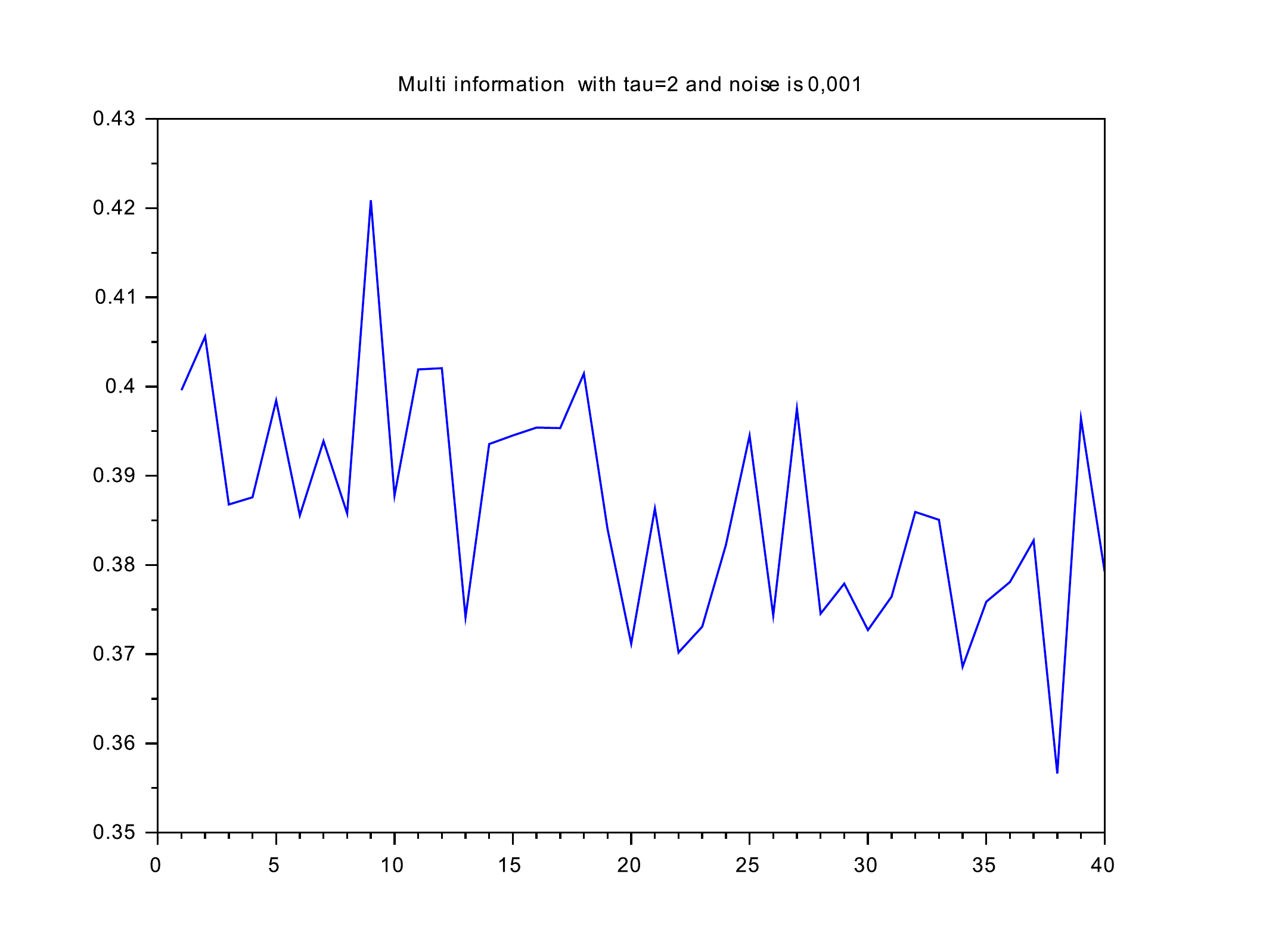,width=.5\textwidth}
 \end{center}
\caption{$\tau$-delayed multi-information $w_i(\tau)$ ($\tau=2$) as a function of
  $i$, for graph $G$  with $n=40$ agents and noise level $\epsilon=0.001$.}
 \label{fig:Multi_Info}
 \end{figure}

The $\tau$-delayed multi-information can be used as a measure of the
influence of opinion of each node $i$ on the vote of the other agents.
For instance, Fig.~\ref{fig:Multi_Info} shows $w_i(\tau=2)$ in a
steady state, where the origin of time is arbitrary. We observe that
some agents $i$ exhibit a more pronounced peak of multi-information
towards the rest of the system, suggesting that the opinion of these
agents may affect the global opinion of all agents. Note that this
results is obtained only by probing the systems, without modifying any
of its components. For this reason, we describe this approach as
``non-intrusive''. The algorithms used throughout this paper to numerically evaluate the delayed mutual- and multi-informations in the voter model example are described in  \ref{Algo1}


\subsection{Intrusive characterisation: forcing}

In this section, we consider another way to measure the influence of
an agent on the system. We call this approach ``intrusive'' as it
implies a perturbation, and no longer just an observation.

To measure the influence of agent $i$, its opinion is forced to a
chosen value, for instance the value 1. As a result the density
(\ref{eq:rho}) of opinions 1 on the system
\begin{equation}
  \rho(t)={1\over n}\sum_{j=1}^n s_j(t)
\end{equation}
can be averaged over a large number $N$ of independent realizations,
to give a quantity $\langle\rho(t)\rangle_i$, where the subscript $i$
indicate which agent has been forced to 1. If $t$ is large enough,
$\langle\rho\rangle_i$ no longer depends on $t$. 

The influence can be measured in a steady state, or from the initial
state where all agents are initialized uniformly to 0 or 1 with
probability 1.

\begin{figure*}[htb]
\begin{center}
\psfig{figure=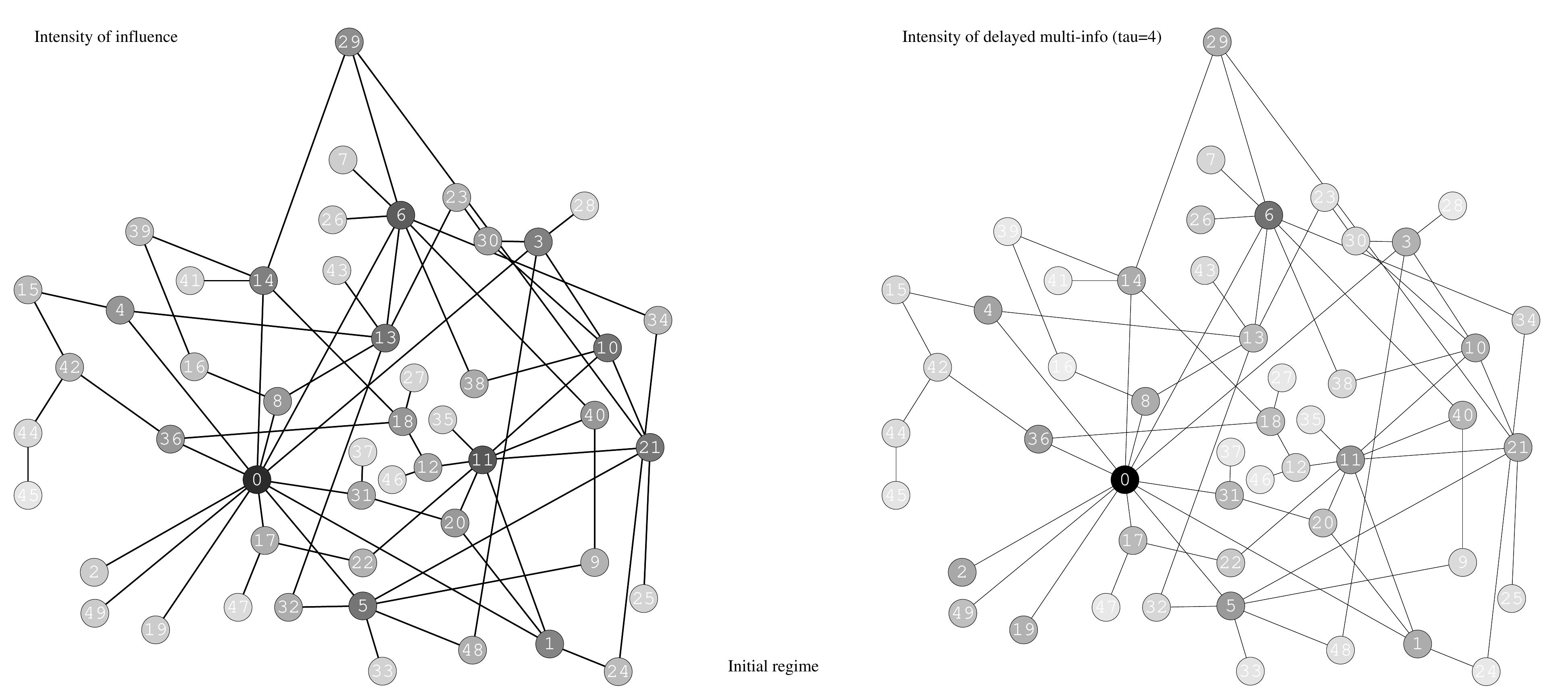,width=.9\textwidth}

\end{center}
\caption{Scale free graph colored as a function of the values of the
  influence (left) and the $\tau$-delay multi-information (right), for
  $\tau=4$. In this case, the multi-information is computed from the
  initial state.}
\label{fig:graph_colored}
\end{figure*}

\begin{figure*}[htb]
\begin{center}
\psfig{figure=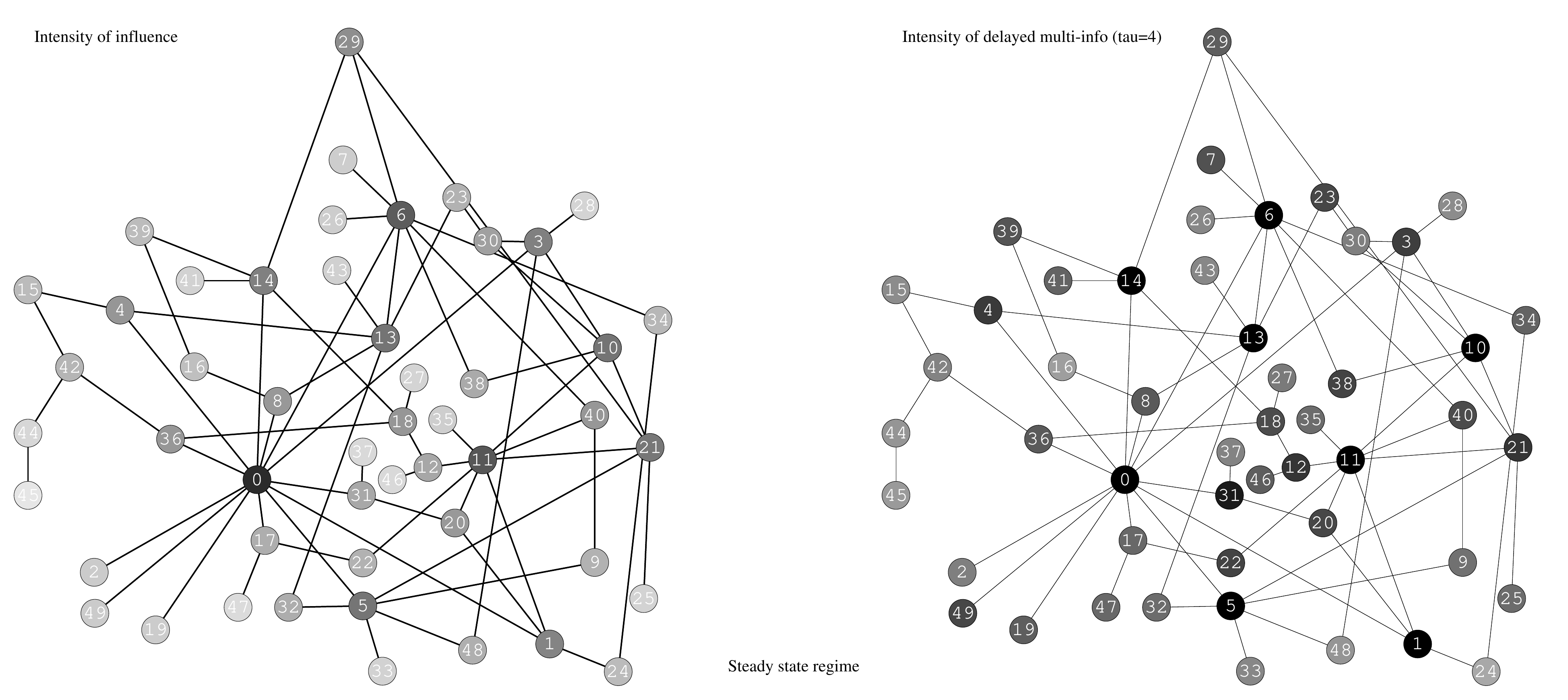,width=.9\textwidth}
\end{center}
\caption{Scale free graph colored as a  function of the values of the
  influence (left) and the $\tau$-delay multi-information (right), for
  $\tau=4$. In this case, the multi-information is computed when the
  system is in a steady state regime.}
\label{fig:graph_colored_steady}
\end{figure*}

The color representations of the graphs (Figures
~\ref{fig:graph_colored} and ~\ref{fig:graph_colored_steady}) show
that the multi-information give  some information about the
controllability and the observability of the system. In the case
the multi-information is calculated from the initial state, these
figures emphasize the link between the multi-information and the
influence of an agent. We can then identify the agents that allow the best
control of the system when their vote is  forced.

The measurement obtained in the steady state for the delayed
multi-information is different from that observed in the transient
regime. Low-impact agents can get a  high multi-information by being a proxi of an influential neighbor. In this case,
the multi-information rather evaluates the observability than the controlability.

\section{The 1D Voter model}\label{sect:1D}

The previous section gave an illustration of the link between
influence defined by intrusive forcing and the influence measured by
observing the time-delayed multi-information. In this section, we
propose an analytical meanfield solution of the voter model, in a
one-dimensional topology. This solution will formally specify the
proposed links. In particular we will introduce  a characteristic
control length.

\subsection{Presentation}

We consider the case of $n$ voters organized along a line so
that voter $i$ looks at voter $i-1$ and itself to take its
decision. Agent $i=0$ has no left neighbor and will have a controlled
dynamics. For instance its opinion will be always 1. The other agents
are initialized randomly in $\{0,1\}$. 

Since agent 1 is looking at agent 0, its next state will likely to be
1. And so on for agent $2, 3, \ldots, n$. Intuitively, we could expect
that the entire system will become 1, due to the control imposed by
agent 1. But noise is changing this conclusion.

\begin{figure}[htb]
 \psfig{figure=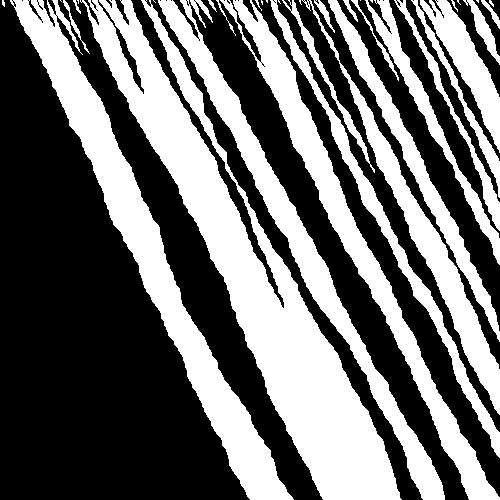,width=.2\textwidth}
 \hfill
 \psfig{figure=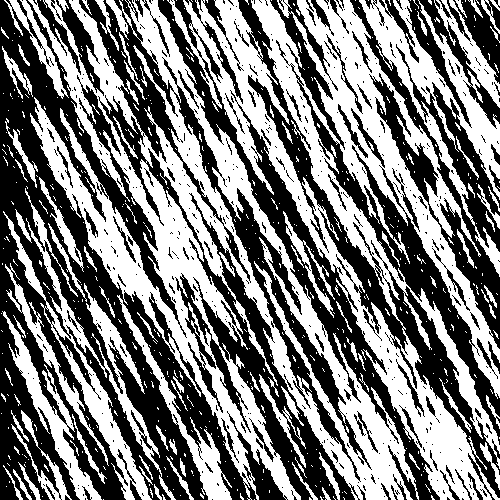,width=.2\textwidth}
 \caption{Space-time diagram of the evolution of the states of a
   $n=500$ agents of a voter model organized in a line. Line $t$ of
   the figure depicts the configuration of the $n$ voter at iteration
   $t$. We can see the first 500 iterations. Left: $\epsilon=0$.
   Right: $\epsilon=0.01$ .}
\label{fig:noise-config}
\end{figure}

\begin{figure*}[htb]
 \psfig{figure=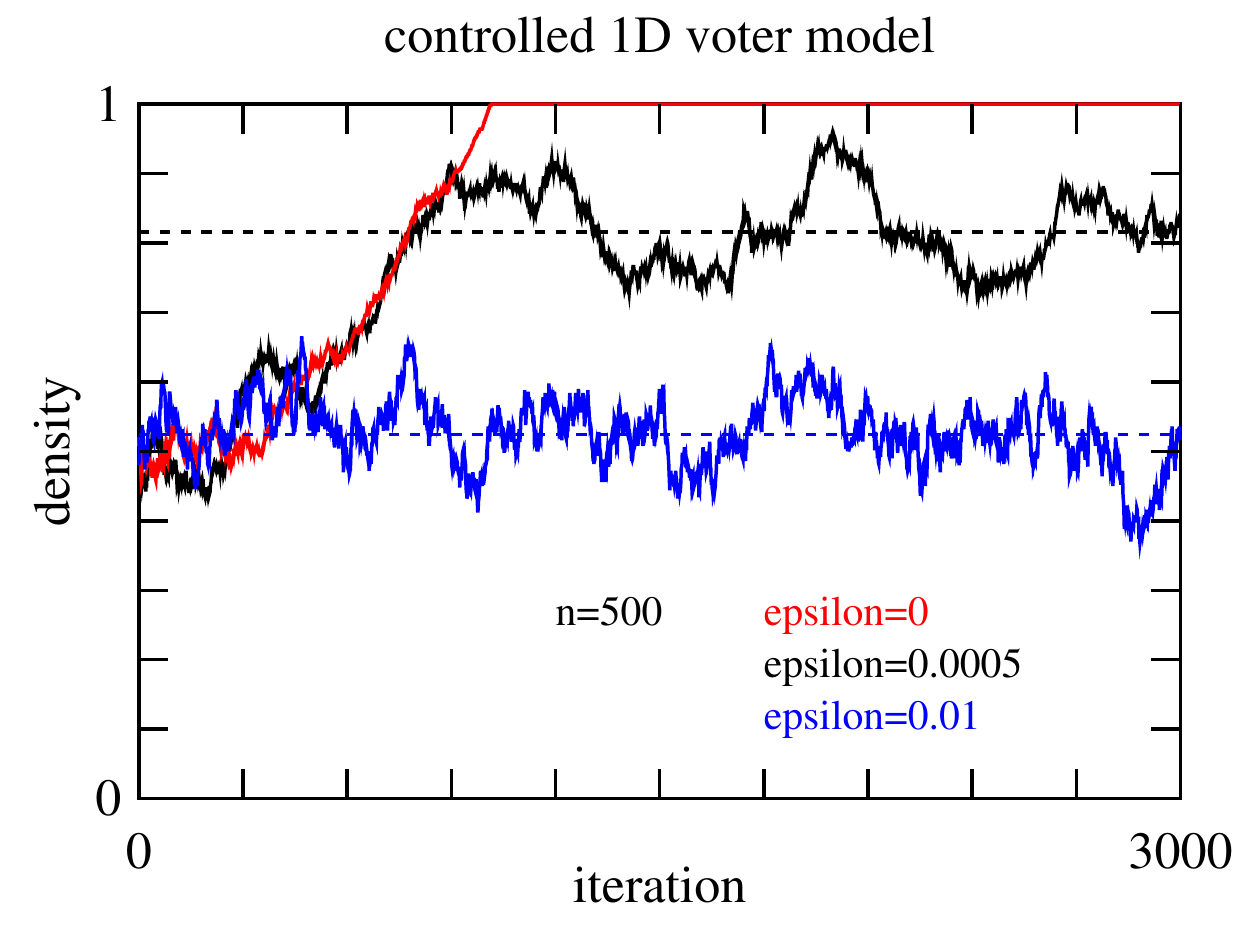,width=.45\textwidth}
 \hfill
 \psfig{figure=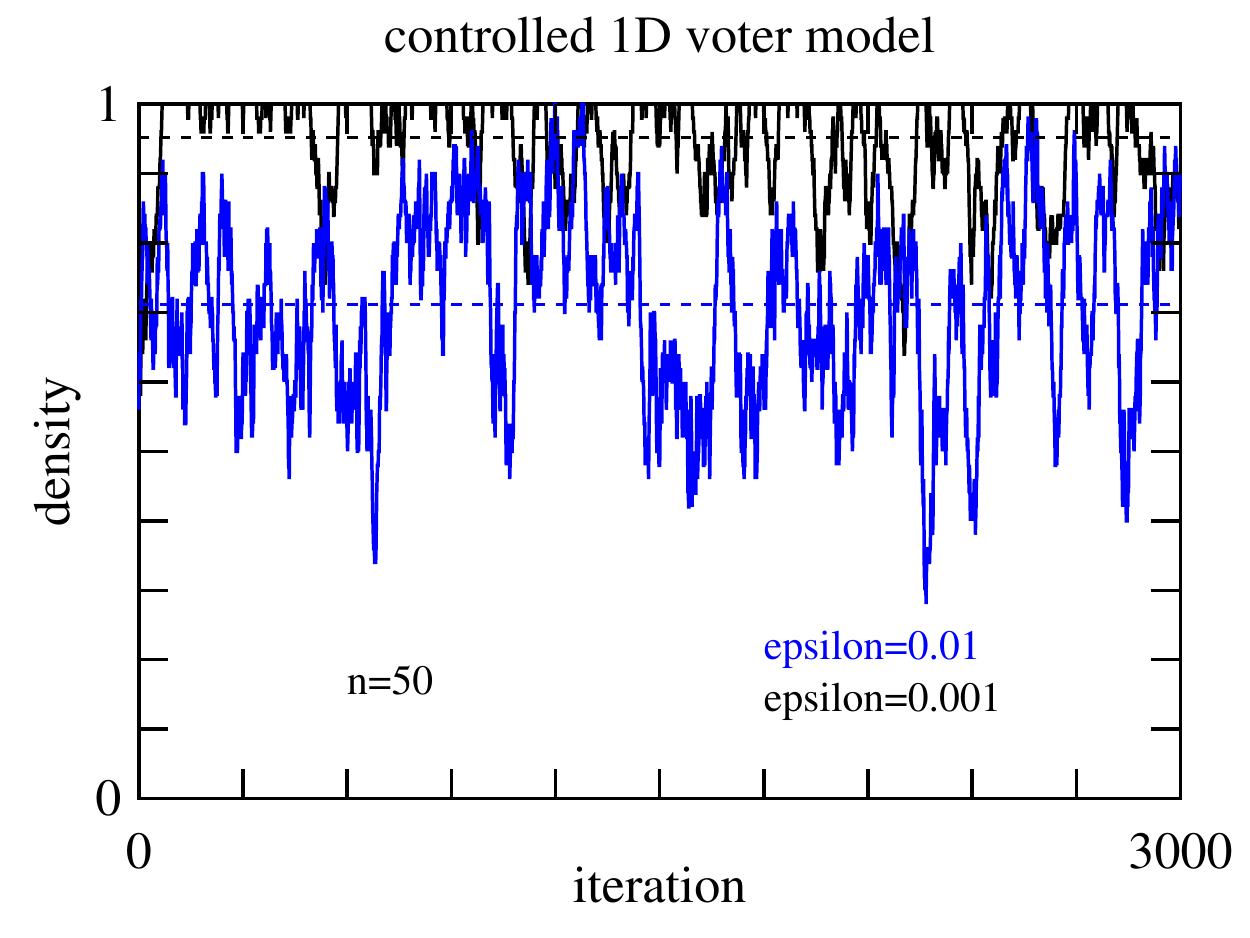,width=.45\textwidth}
 \caption{Density of agents with value 1 as a function of time, for
   different noise levels, and two different system sizes, $n=500$ and
   $n=50$. The dashed lines are the predictions of the meanfield
   analytical approach, see eq.~(\ref{eq:average-asymptotic-value}).}
\label{fig:density-over-time}
\end{figure*}

If $p_i(t)$ is the probability that agent $i$ is 1 at time $t$, we can write
the equation
\begin{equation}
p_i(t+1)=p_i(t)W_{1\to1}(t) + (1-p_i(t))W_{0\to 1}(t)
\end{equation}
where $W_{a\to b}$ is the probability that the state evolves from $a$ to $b$.
In  a meanfield approximation, we can write, 
\begin{eqnarray}
W_{0\to 1} &=& p_{i-1}(t)f(1/2) + (1-p_{i-1}(t))f(0)
\nonumber\\
W_{1\to 1} &=& p_{i-1}(t)f(1) + (1-p_{i-1}(t))f(1/2)
\nonumber\\
\end{eqnarray}
Before attempting to solve the above system analytically, we can
observe its behavior numerically. We can see on
Fig.~\ref{fig:noise-config} that is the noise is small ($\epsilon=0$),
the entire system is indeed controlled by the left-most agent whose
state is always 1. But if the noise is increased ($\epsilon=0.01$) the
control is not effective anymore. There is a critical noise
$\epsilon=\epsilon_c(n)$ below which a system of size $n$ can be controlled by the
first node, and above which the influence of the driving node is
diluted by the noise.
Figure~\ref{fig:density-over-time} shows the density of agents with
opinion 1, as a function of time, for different intensities of noise,
$\epsilon$.  We observe in this figure the effect of the system
size. For smaller systems, the effect of controlling agent $i=1$ is
more effective than for larger $n$.

\subsection{Probability distribution of the system}

We can determine the probability distribution in the case of the linear voter
model. We have
\begin{align*}
p_i(t+1) & = p_i(t) W_{1 \to 1}(t) + (1-p_i(t)) W_{0 \to 1}(t)\\
&= p_i(t)  ( W_{ 1 \to 1}(t) - W_{ 0 \to 1} (t)) + W_{0 \to 1} (t)   
\end{align*}
With
\begin{align*}
W_{ 1 \to 1} (t)& = p_{i-1}(t) f(1) + (1-p_{i-1}(t))f(1/2)\\
         &= p_{i-1}(t) (1 - \epsilon)  + (1-p_{i-1}(t)) \dfrac{1}{2} \\
\end{align*}
and         
\begin{align*}
W_{ 0 \to 1}(t) &= p_{i-1}(t) f(1/2) + (1-p_{i-1}(t))f(0) \\
  &= p_{i-1}(t) \dfrac{1}{2} + (1-p_{i-1}(t))\epsilon
\end{align*}
we obtain 
\begin{equation}
p_i(t+1) = \left( \dfrac{1}{2} - \epsilon \right) p_i(t) + \left( \dfrac{1}{2} - \epsilon \right) p_{i-1}(t) + \epsilon \label{rel_distribution}
\end{equation}
As $p_0(t) = 1$, we obtain
\begin{equation}
\ p_1(t+1) = \left( \dfrac{1}{2} - \epsilon \right) p_1(t) + \dfrac{1}{2} 
\end{equation}

Let $P(t)$ be the vector of probability defined by  
$$P(t) = \begin{pmatrix}
p_1(t) \\ 
p_2(t) \\ 
\vdots \\ 
p_n(t)
\end{pmatrix} $$
With this notation, the system can be expressed in  a matrix form
\begin{equation}
 P(t+1) =A P(t) + B
\label{eq:matrix-form}
 \end{equation} 
with
\[ A = ( \dfrac{1}{2} - \epsilon ) \begin{pmatrix}
1 & 0 & \ldots & \ldots  & 0 \\ 
1 & 1 & 0 & \ddots & \vdots \\ 
0 & 1 &1 & \ddots & 0 \\ 
\vdots &  & \ddots & \ddots &0\\ 
0 & \ldots & 0 & 1 & 1
\end{pmatrix}
\]
and
\[ B= \begin{pmatrix}
1/2 \\ 
\epsilon \\ 
\vdots \\ 
\epsilon
\end{pmatrix}
\]
This equation can be solved recursively and gives
\begin{equation}
P(t) = A^t P(0) + \big( \sum_{j=0}^{t-1} A^j \big) B
\label{eq_Jordan}
\end{equation}
The explicit forms for power matrices $A^j$ are given in \ref{Appendix_Jordan}.

\subsection{Stationnary system}

Let us write
\[ \Pi = \begin{pmatrix}
\pi_1 \\ 
\pi_2 \\ 
\vdots \\ 
\pi_n
\end{pmatrix}
\]
the stationary distribution.  From relation~(\ref{eq:matrix-form}), it obeys

\begin{equation}
  A \Pi +B = \Pi \Leftrightarrow  \begin{cases} (\dfrac{1}{2} + \epsilon ) \pi_1 = \dfrac{1}{2}\\
    \forall i \in   \llbracket 2;n \rrbracket,\ \pi_i = \dfrac{1 - 2 \epsilon}{1 + 2 \epsilon} \pi_{i-1} + \dfrac{ 2 \epsilon}{1 + 2 \epsilon} \end{cases}
\end{equation}
It is an arithmetico-geometric sequence which can be solved for all agents $i$ as
  \begin{equation}
   \begin{cases}
   \pi_1 = \dfrac{1}{1+2 \epsilon}\\
  \pi_i = \dfrac{1}{2} + \Big( \dfrac{1 - 2 \epsilon}{1 + 2 \epsilon} \Big)^{i-1} (\pi_1 - \dfrac{1}{2})
  \end{cases}  \label{eq:pi-1}
  \end{equation}
with 
\[ \pi_1-{1\over2}={1\over2}\left({1-2\epsilon\over 1+2\epsilon} \right)\]

Further, we can write eq.(~\ref{eq:pi-1}) as 
\begin{eqnarray}
  \pi_i&=&{1\over 2}\left[1 + \left({1-2\epsilon\over 1+2\epsilon}\right)^i\right]
  \nonumber\\
  &=&{1\over 2} + {1\over2}\exp\left[-i\ln\left({1+2\epsilon\over 1-2\epsilon}\right)\right]
  \nonumber\\
  &=&{1\over 2} + {1\over2}\exp\left[-{i\over\ell_c}\right]
\label{eq:pi-2}
\end{eqnarray}
where $\ell_c$ is defined as 
\begin{equation}
\ell_c={1\over\ln\left({1+2\epsilon\over 1-2\epsilon}\right)}
\label{eq:control-length}
\end{equation}
and referred to as the {\it control length} as it gives a value for
$i$ above which the exponential falls quickly to zero.  It is a characteristic distance from the controlled agent where its inluence starts to fade.

\begin{figure}
\centering
\psfig{figure=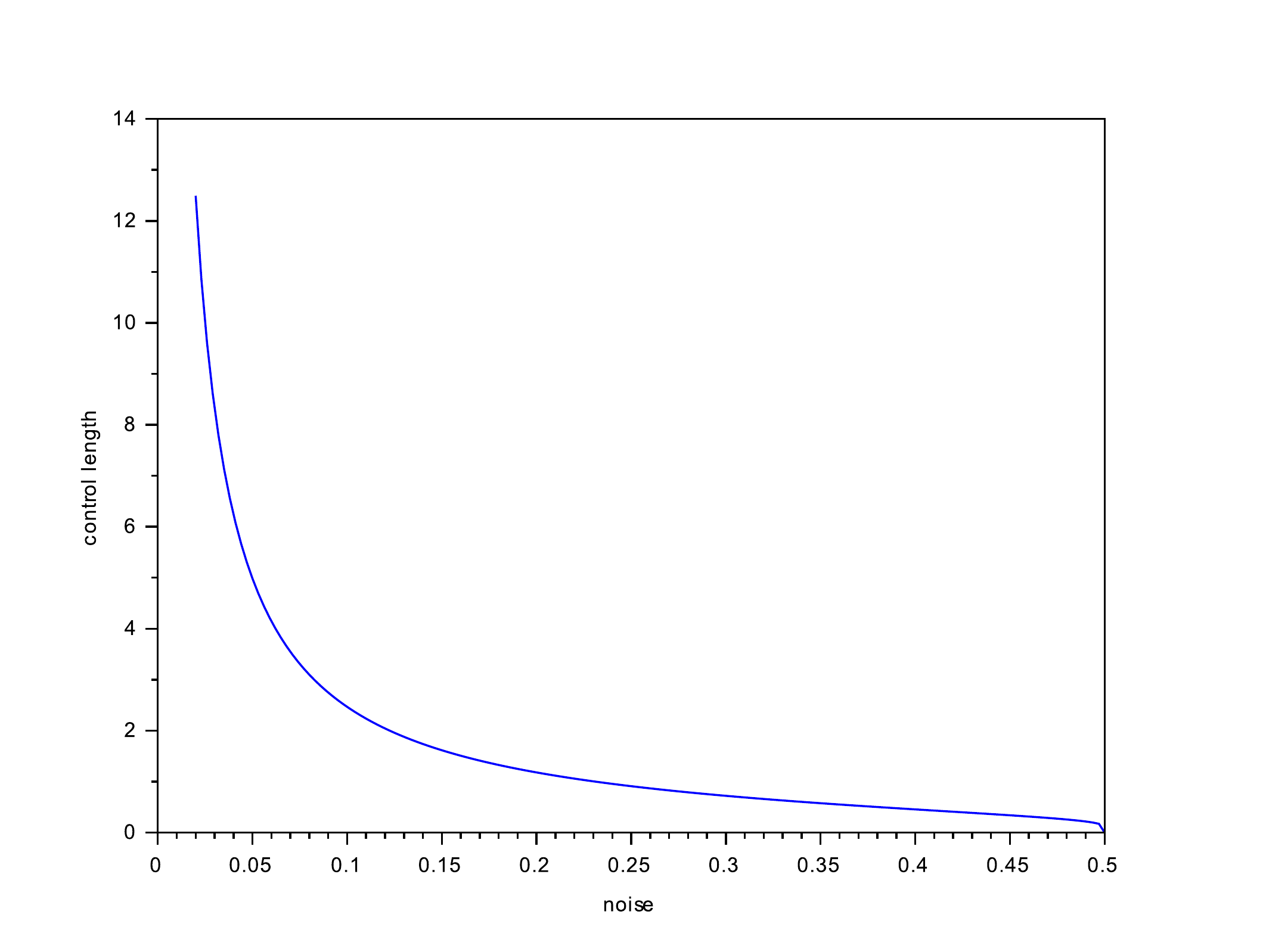,width=.4\textwidth}
\caption{Control  length $\ell_c$ as a function of the noise 
  $\epsilon$, according to eq.~(\ref{eq:control-length}).}
\label{fig:control-length}
\end{figure}

We see that, when $\epsilon$ approaches $1/2$, the length of control
$\ell_c$ converges to $0$, which corresponds to a total loss of the
controlability of the system. Figure~\ref{fig:control-length} shows that
$\ell_c$ decreases very quickly to $0$ when $\epsilon$ increases to
1/2.

\subsection{Average vote of the system}

In the case of a stationnary system, we can calculate the average
density of agents with vote $1$.
\[ S={1\over n}\sum_{i=1}^{n} \pi_i \]
with $n$ is the number of free agents.  

According to~(\ref{eq:pi-2}), we have 
\begin{equation}
S={1\over n}\left[ {n\over2} + 
     {1\over2}\sum_{i=1}^n\left({1-2\epsilon\over 1+2\epsilon}\right)^i \right]
\end{equation}

When $\epsilon\ne 0$, we have
\begin{eqnarray}
\sum_{i=1}^{n}\left({1-2\epsilon\over 1+2\epsilon}\right)^i
&=&
\left({1-2\epsilon\over 1+2\epsilon}\right) \sum_{i=0}^{n-1}\left({1-2\epsilon\over 1+2\epsilon}\right)^i\\
&=&\left({1-2\epsilon\over 1+2\epsilon}\right) {1-\left({1-2\epsilon\over 1+2\epsilon}\right)^{n}
    \over
    1- \left({1-2\epsilon\over 1+2\epsilon}\right)} 
\nonumber\\
\end{eqnarray}
and we obtain 
\begin{eqnarray}
S&=&{1\over 2}+{1\over 2n}\left[ \left({1-2\epsilon\over 1+2\epsilon}\right) {1-\left({1-2\epsilon\over 1+2\epsilon}\right)^{n}
    \over
    1- \left({1-2\epsilon\over 1+2\epsilon}\right)}  \right]
\nonumber\\
&=&{1\over 2}+{1\over 2n}\left({1-2\epsilon\over 4\epsilon}\right) \left[  1-\left({1-2\epsilon\over 1+2\epsilon}\right)^{n}   \right]
\nonumber\\
\label{eq:average-asymptotic-value}
\end{eqnarray}
In Figure~\ref{fig:density-over-time}, we see that the simulations are
in agreement with this theoretical result.

\subsection{Delayed mutual information}

In this section we will compute the influence of an agent based on the
$\tau$-delayed mutual information, $w_{i,j}(\tau)$, between agents $i$
and $j$, as defined in eq.~(\ref{info_mutual}). These values are
obtained by a sampling of the simulation of the 1D voter model, with
$n=50$ agents.  Measurements are performed when the system has reached
a stationnary state, that is after $t$ iterations such that all the
probabilities $A^t P(0)$ are smaller than a certain threshold. In our
case, we take the threshold at $10^{-4}$.

\begin{figure*}
 \psfig{figure=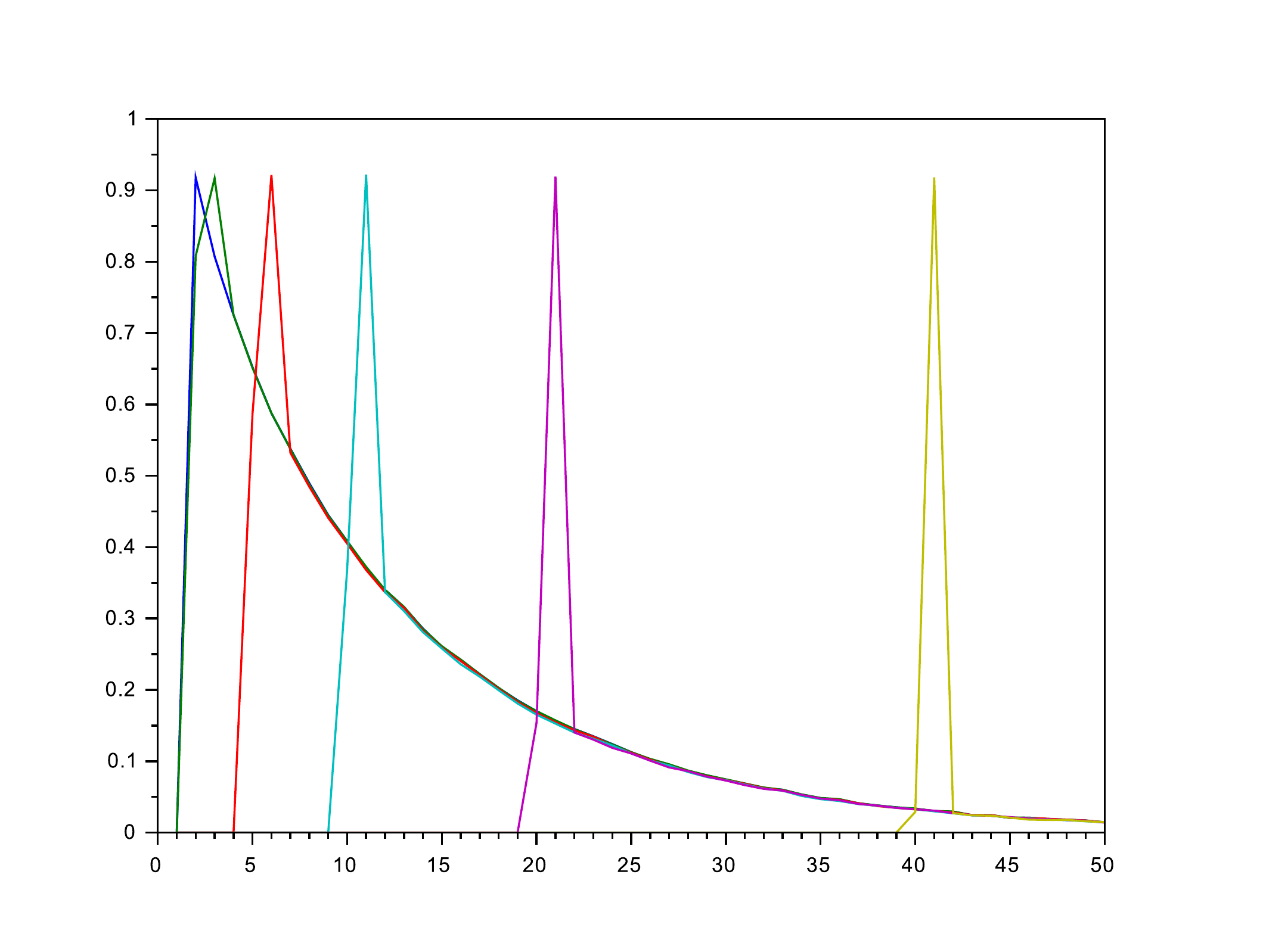,width=.3\textwidth}
\hfill
  \psfig{figure=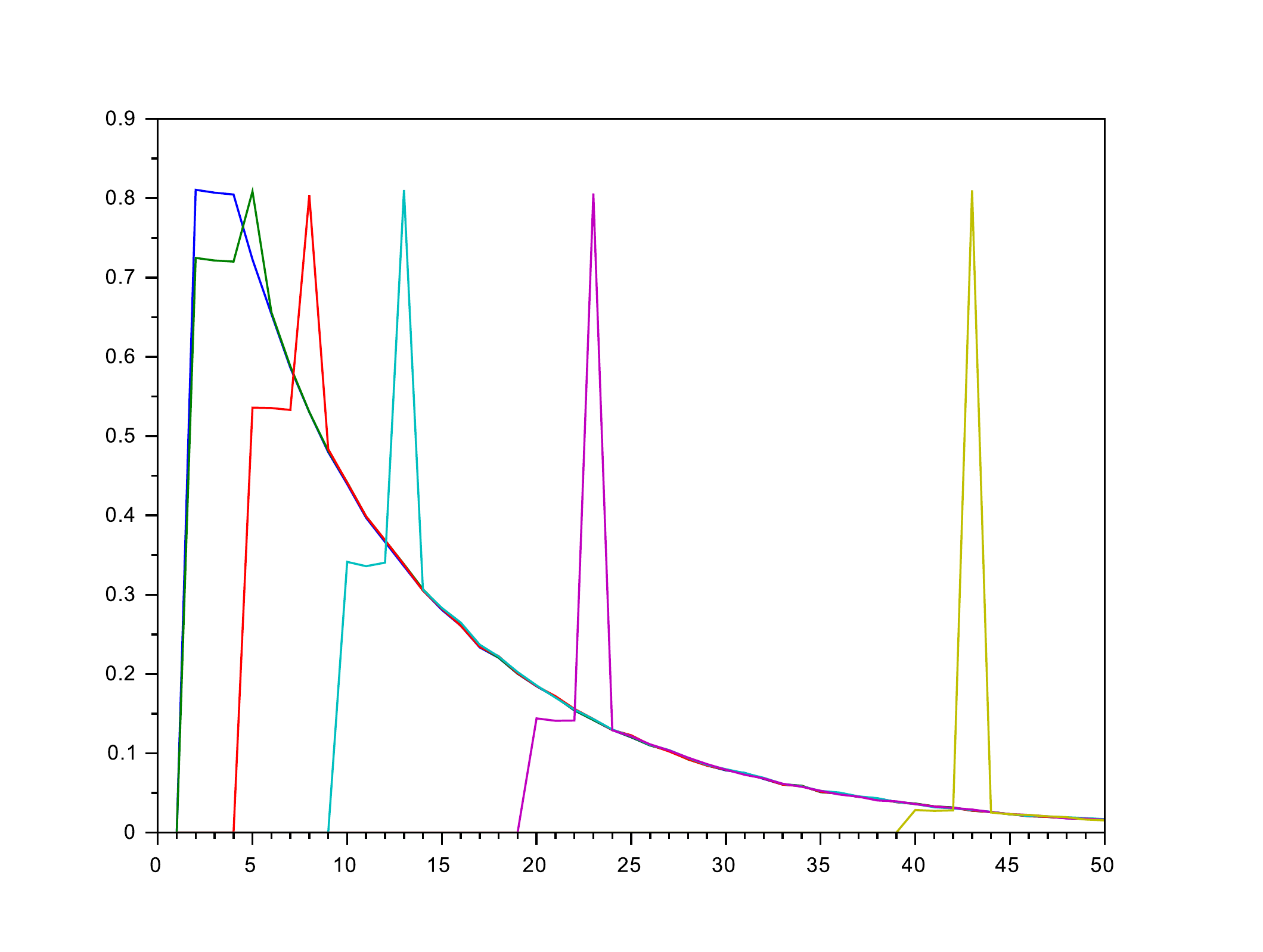,width=.3\textwidth}
 \hfill
    \psfig{figure=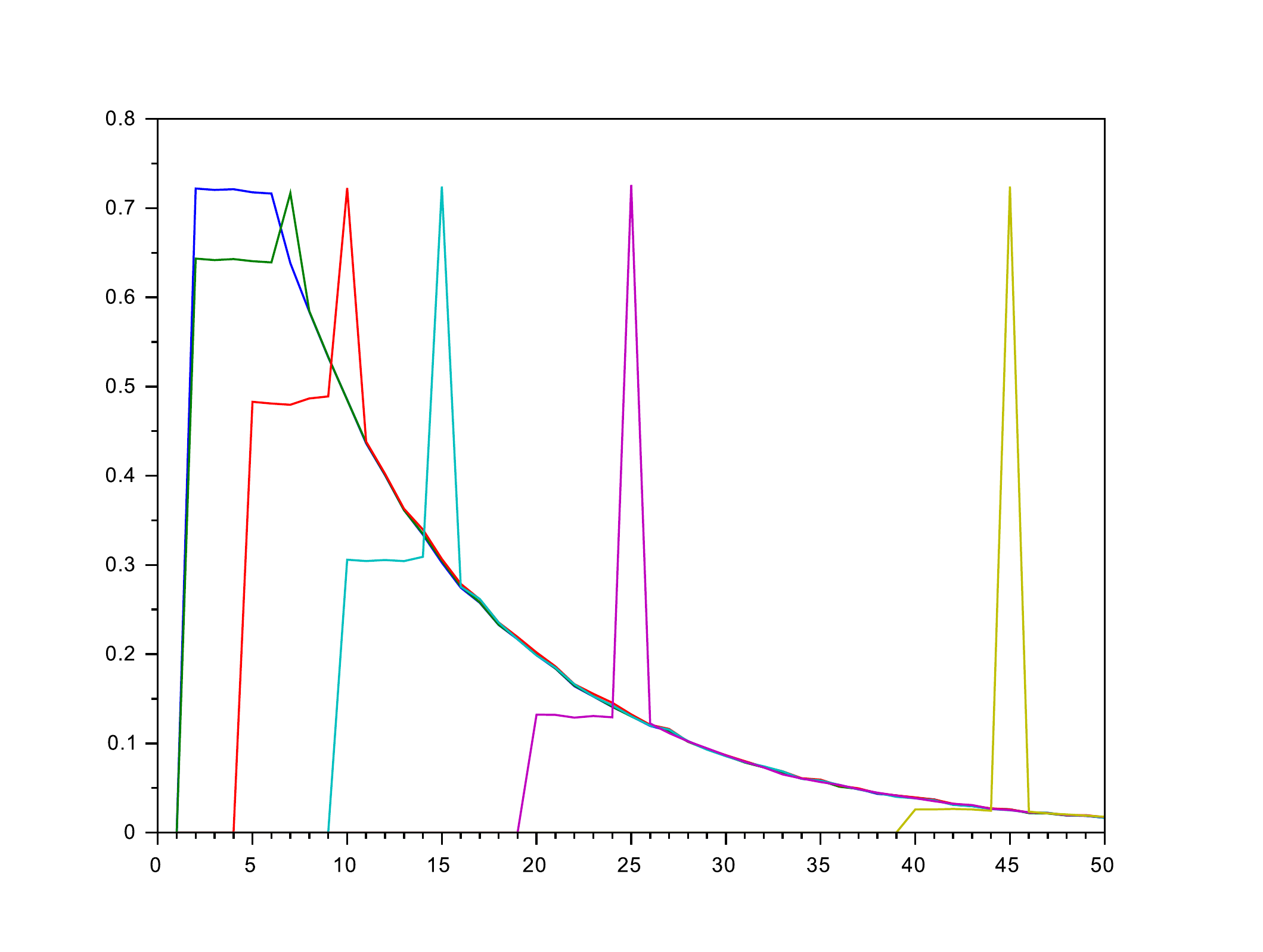,width=.3\textwidth}
 \caption{Delayed mutual information, $w_{i,j}(\tau)$, as a function
   of agent $j$, for different values of $i$. The different curves
   correspond to $i=1,\ 2,\ 5,\ 10,\ 20$ et $40$, from left to right,
   respectively. The vote of agent $i=0$ is forced to $1$ and the
   noise is $\epsilon=0.01$.  The delay is $\tau =1$ (left panel), $\tau =3$
   (middle panel) and $\tau =5$ (right panel).}
\label{fig:info_mutuelle1}
\end{figure*}

In Fig.~\ref{fig:info_mutuelle1}, we notice that the mutual
information $w_{i,j} (\tau)$ is zero if $j<i$, has a plateau for
$j<i+\tau$, shows a peak for $j=i+\tau$, and decreases for
$j>i+\tau$. This observation reflects the fact that agent $i$ can only
influence agents on its right as the voting decision of an agent is
based on the state of its left neighbor. The plateau shows the
influence of the past $j-i$ iterations. The influence of $i$ over $j$
is maximum for $j=i+\tau$ as it takes $\tau$ iterations for the vote
of $i$ to travel from $i$ to $j$. For $j>i+\tau$ the influence is due
to the steady state regime.

\begin{figure*}[htb]
 \psfig{figure=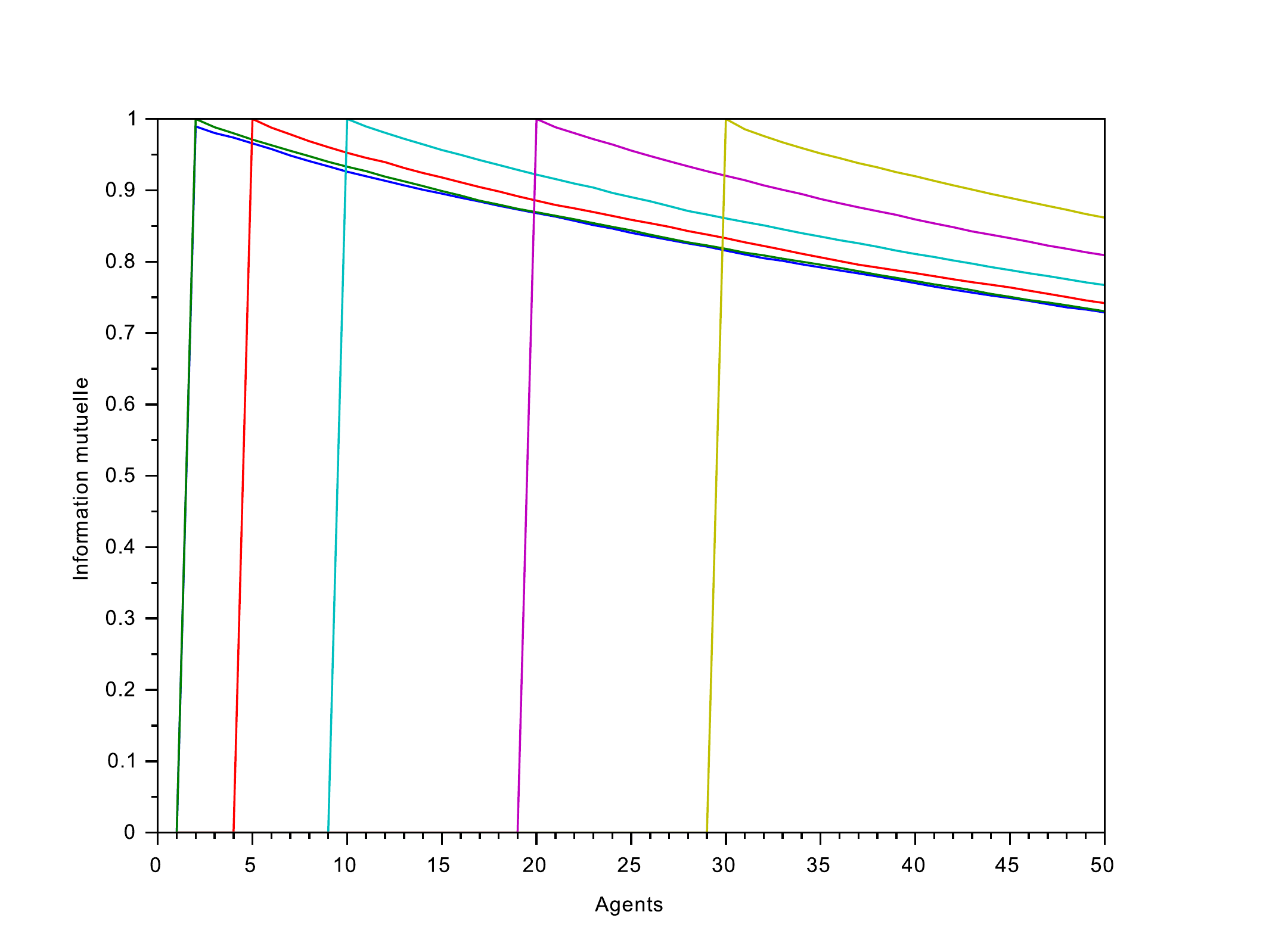,width=.3\textwidth}
\hfill
  \psfig{figure=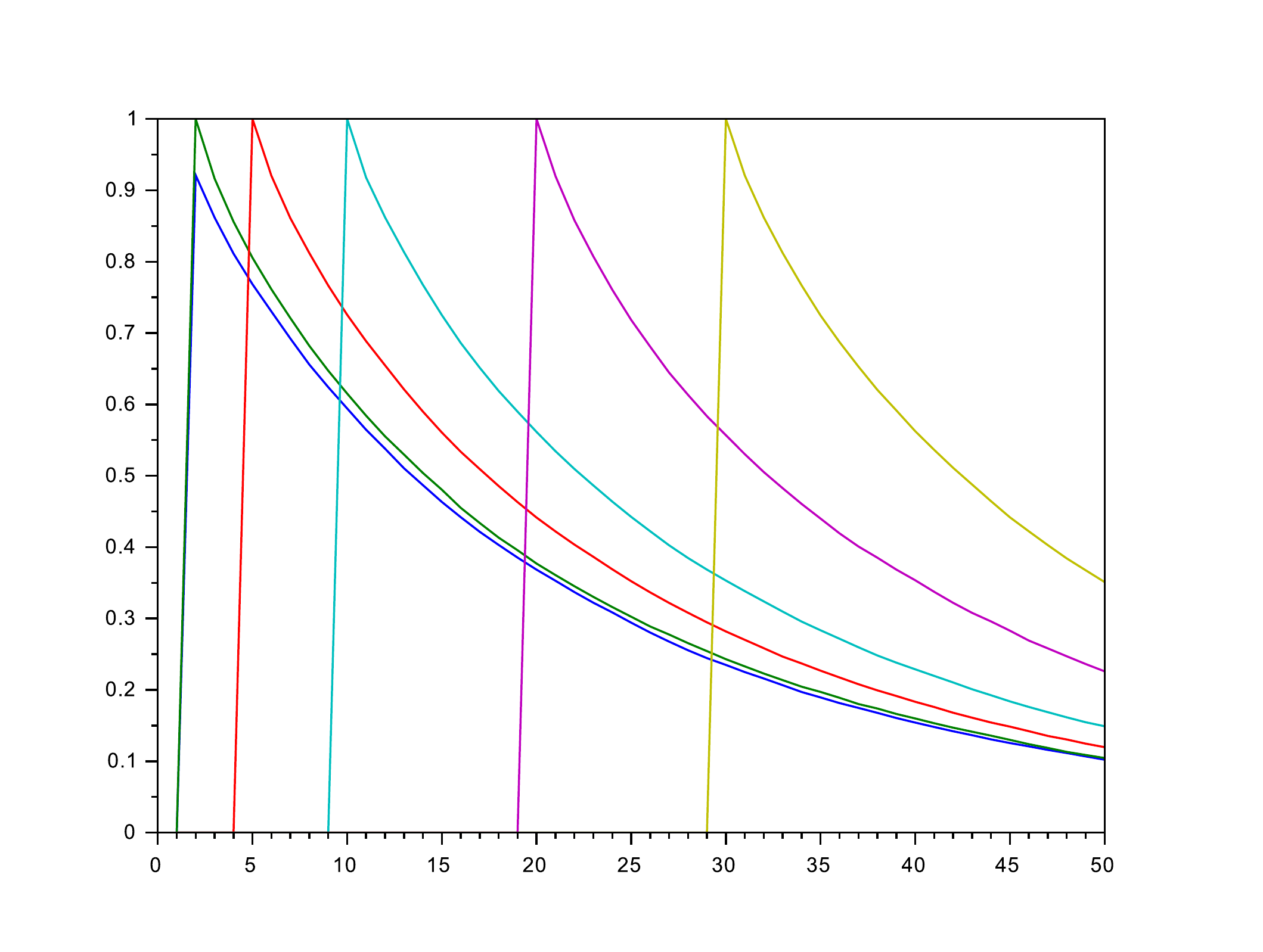,width=.3\textwidth}
  \hfill
  \psfig{figure=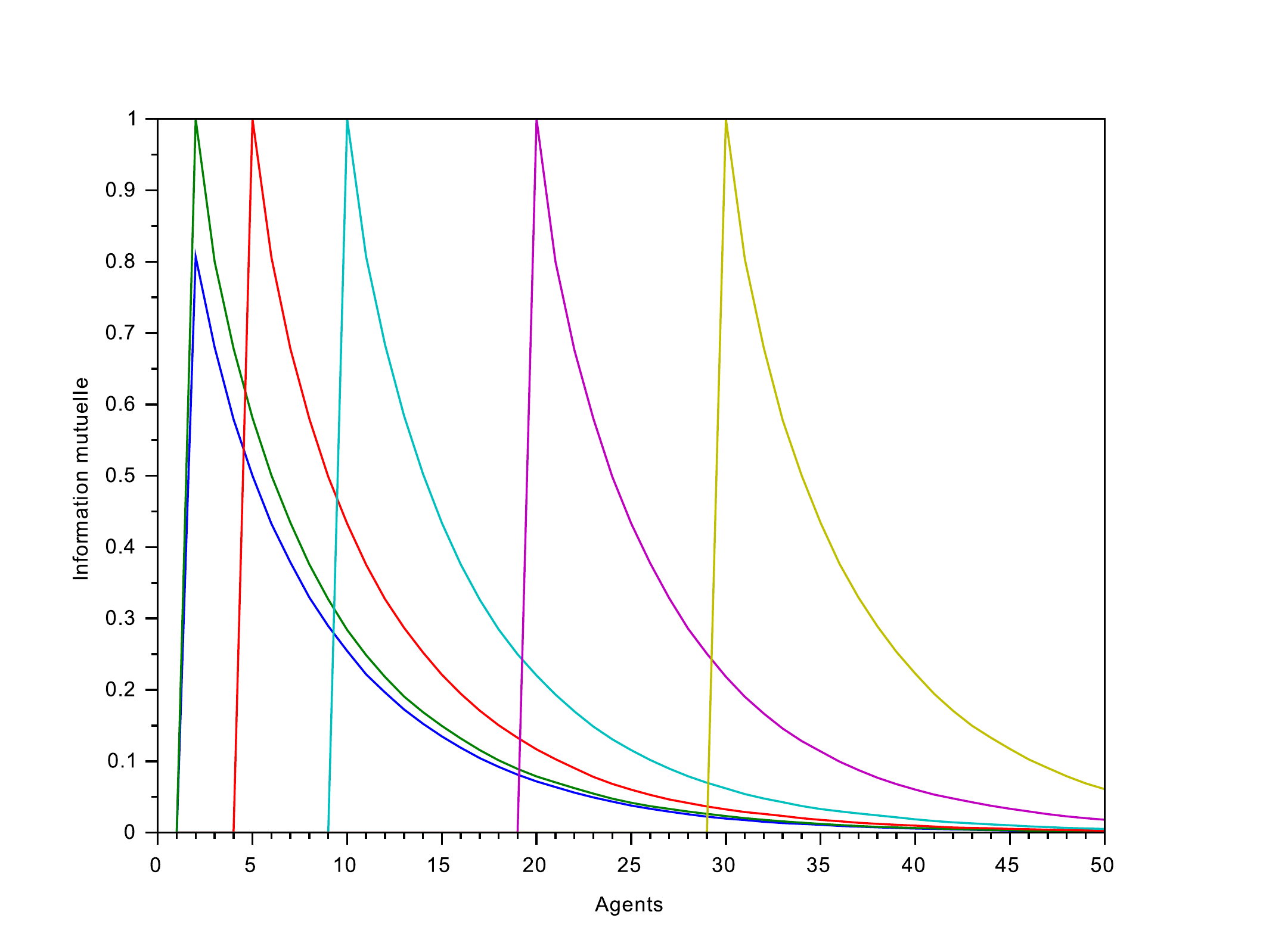,width=.3\textwidth}
 
 \caption{Delayed mutual information $w_{i,j}(j-i)$ as a function of
   $j$, for agents $i=1,\ 2\,\ 5,\ 10,\ 20$ and $30$ (curves from left
   to right, respectively). The vote of agent $i=0$ is forced to
   $1$ and the noise is $\epsilon=0,001$ (left panel), 
   $\epsilon=0,01$ (middle panel) and $\epsilon=0,05$ (right panel).}
\label{fig:info_mutuelle2}
\end{figure*}

In Fig.~\ref{fig:info_mutuelle2} we consider the behavior of
$w_{i,j}(j-i)$. It suggests the following  relationship
\begin{equation}
  \forall j>i, w_{i,j}(j-i) = \alpha_i \exp\left[- \lambda_i (j-i)\right]
  \label{eq:wij}
\end{equation}
where $\alpha_i$ and $\lambda_i$ depend on the noise level, $\epsilon$.

 \begin{figure}[htb]
 \psfig{figure=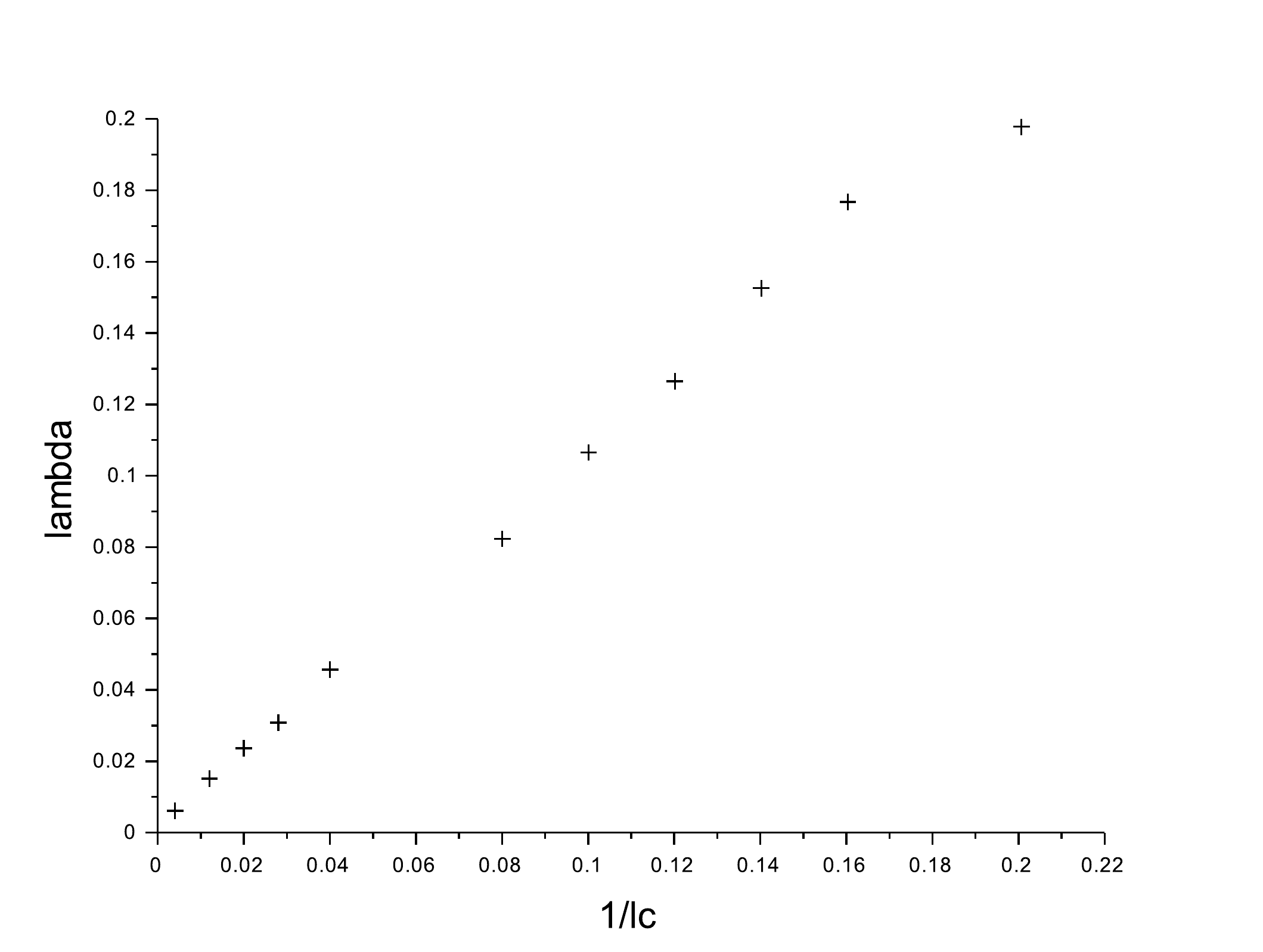,width=0.5\textwidth}
 \caption{$\lambda$ as a function of 1/ $\ell_c$}
\label{fig:lambda_lc}
\end{figure}

 \begin{figure*}[htb]
\psfig{figure=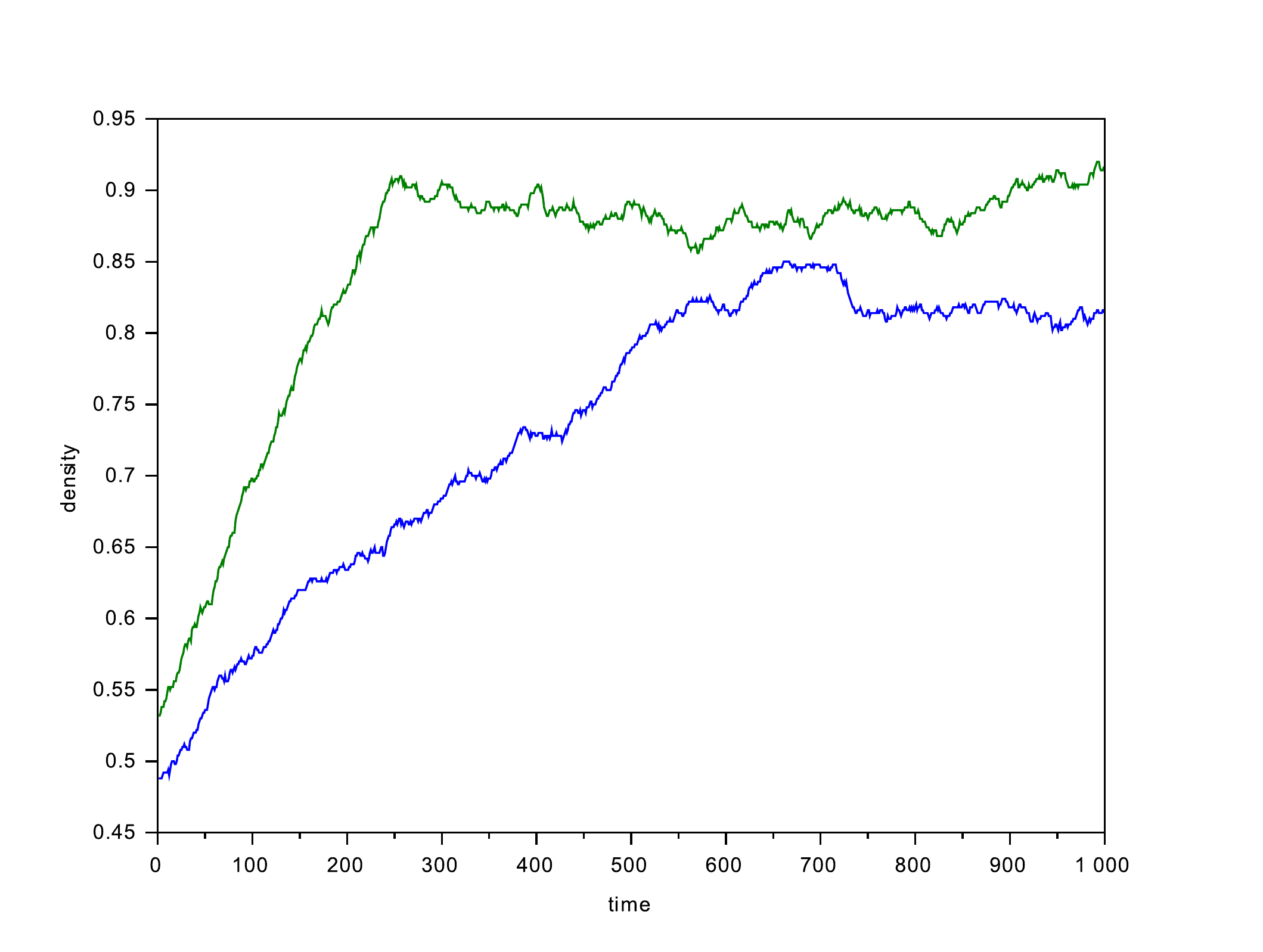,width=.3\textwidth}
\hfill
 \psfig{figure=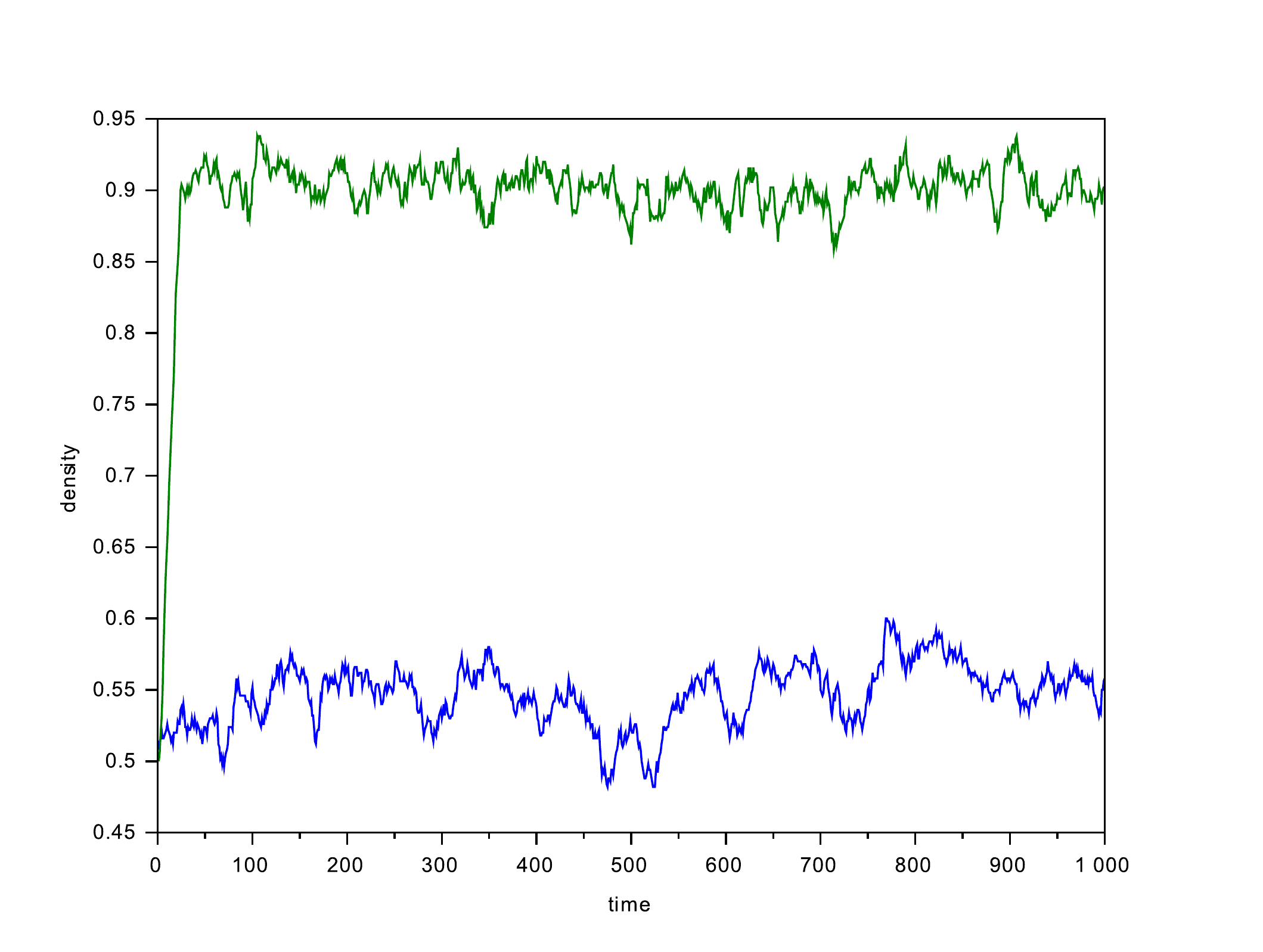,width=.3\textwidth}
\hfill
 \psfig{figure=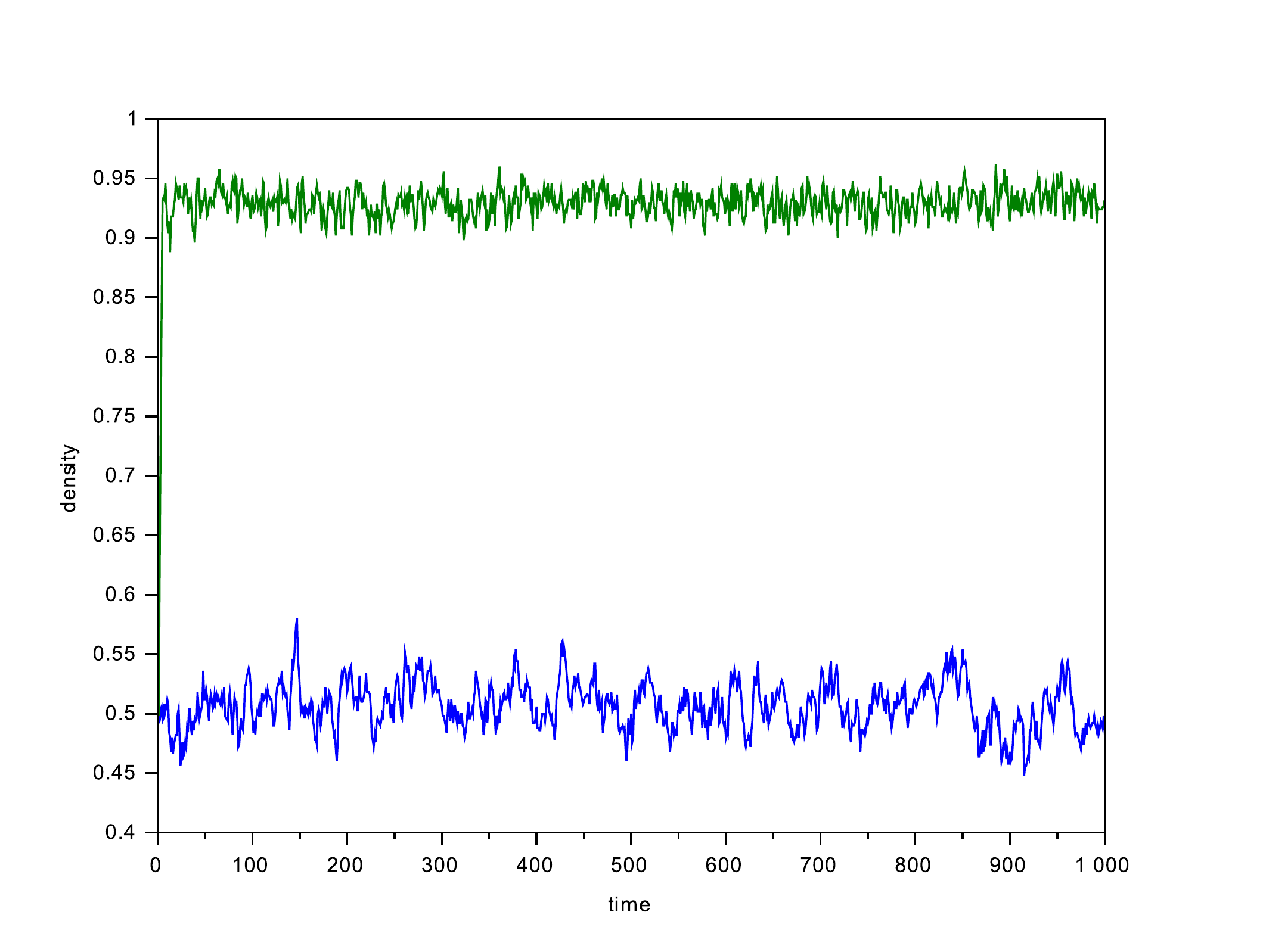,width=.3\textwidth}
 \caption{Evolution of the density of vote $1$ during the time when only the vote of the agent 1 is forced to 1 (blues curves) and when every vote of the agents $j \propto \lfloor \dfrac{1}{ \lambda} \rfloor$ is forced to 1 (greens curves). 
  left  $\epsilon=0.001$. middle :$\epsilon=0.01$,  right: $\epsilon=0.05$ .}
\label{fig:tauxlambda}
\end{figure*}

The coefficients of correlation between $\ln(w_{i,j}(j-i))$ and $j$,
for different values of the noise are found to be between $-1$ and
$-0.99$, thus confirming the relation proposed in
eq.~(\ref{eq:wij}). The value of $\alpha_i$ and $\lambda_i$ can be
determined with a least squares method.

Consequently, the value of the delayed mutual information $w_{i,j}(j-i)$
decreases quickly as $j$ departs from $i$. This reflects the difficulty
to control agent $j$ from agent $i$.


This interpretation is confirmed by Fig.~\ref{fig:lambda_lc} which
shows the relation between the values of $\lambda\equiv\lambda_1$ and
the control length $\ell_c$ defined in
eq.~(\ref{eq:control-length}). Each point in this figure corresponds
to a different value of the noise. The relation can be fitted by
\begin{equation}
\lambda(\epsilon) = a \dfrac{1}{ \ell_c(\epsilon)} + b
\label{eq:lc_lambda}
\end{equation}
with $a = 0.973$ and $b=-0.003$, independent of the value of
$\epsilon$.  The coefficient of correlation is $0.997$, in agreement
with the proposed linear link between $\lambda$ and $1/\ell_c$.

\subsection{Control of the density of vote $1$}

The previous section suggests that the influence of an agent decreases
exponentially with the distance to others, with a characteristic
length which decreases as the noise increases. This result follows both
from studying an intrusive action on the system, or by simply
observing it. In this section we exploit this result to find a
strategy to control the full system by acting on more than one
agent. In practice we consider the situation where agents $i=kd$ are
forced to vote 1, where $k\in\{0,1,2,\ldots\}$ and $d$ is given by the
control length $\lfloor \ell_c \rfloor $ or $\lfloor \frac{1}{ \lambda_1} \rfloor$.

Fig.~\ref{fig:tauxlambda} shows the simulation results for $d$ chosen
as $\lfloor \frac{1}{ \lambda_1} \rfloor$. The density of agents voting
$1$ increases significantly. The quantity $n\lambda_1$ is the number of
controlled agents and is a good indicator to evaluate the cost to
control the system.

\subsection{Noise and information capacity}

In the previous section, by evaluating the mutual information, we
found that the cost of control increased greatly when the noise
increases. This result can be related to the notion of {\it capacity},
as defined in the standard theory of information.  In the linear voter model,
agent $i+1$ can be considered as a channel of communication where the
input message is the vote of agent $i$ at the time $t$ and the
output message is the vote of agent $i+2$ at he time $t+2$.

The information channel capacity $C_2$ is defined as (see
(\cite{cover2012})
\begin{equation}
\displaystyle C_{2} = \sup_{\mathbb{P}_{X_i}} I(X_i(t),X_{i+2}(t+2)) = \sup_{\beta \in [0;1]}  I(X_i(t),X_{i+2}(t+2))
\end{equation}
With $\begin{pmatrix}
\mathbb{P} (X_{i}(t) = 1) \\ 
\mathbb{P} (X_{i}(t) = 0)
\end{pmatrix} = \begin{pmatrix}
\beta \\ 
1- \beta
\end{pmatrix}$,
we obtain
$$\begin{cases} \mathbb{P} (X_{i+1}(t+1) = 1) = (1- \epsilon ) \mathbb{P} (X_i (t) = 1) + \epsilon \mathbb{P} (X_{i}(t) = 0)\\
\mathbb{P} (X_{i+1}(t+1) = 0) = \epsilon \mathbb{P} (X_i (t) = 1) + (1- \epsilon ) \mathbb{P} (X_{i}(t) = 0)
\end{cases}$$
which we write as
$$\begin{pmatrix}
\mathbb{P} (X_{i+1}(t+1) = 1) \\ 
\mathbb{P} (X_{i+1}(t+1) = 0)
\end{pmatrix}  = A \begin{pmatrix}
\beta \\ 
1- \beta
\end{pmatrix} \text{ où } A = \begin{pmatrix}
1- \epsilon & \epsilon \\ 
\epsilon & 1-\epsilon
\end{pmatrix} $$
Therefore,
$$\begin{pmatrix}
\mathbb{P} (X_{i+2}(t+1) = 1) \\ 
\mathbb{P} (X_{i+2}(t+1) = 0)
\end{pmatrix}  = A^2 \begin{pmatrix}
\beta \\ 
1- \beta
\end{pmatrix}$$
where
\begin{eqnarray}
A^2
&=&
\begin{pmatrix}
(1- \epsilon)^2 + \epsilon^2 & 2\epsilon (1- \epsilon) \\ 
2\epsilon (1- \epsilon) & (1- \epsilon)^2 + \epsilon^2
\end{pmatrix} \\
&=&\begin{pmatrix}
1-2\epsilon (1- \epsilon) & 2\epsilon (1- \epsilon) \\ 
2\epsilon (1- \epsilon) & 1-2\epsilon (1- \epsilon)
\end{pmatrix}\\
\end{eqnarray}
This corresponds to a binary symmetric channel, with a probability of error
\[ p_e=2\epsilon (1- \epsilon)\].
We know that the capacity of a binary symmetric channel with a probabiliy of error $p_e$ is
\[ C= 1- H_2(p_e) \text{ with } H_2(p_e) = - p_e \log_2(p_e) - (1-p_e) \log_2 (1-p_e)\]
Therefore
\begin{equation}
C_{2} = 1- H_2 (2\epsilon (1- \epsilon)) 
\end{equation}

Now, we consider all agents from $i+1$ to $i+m-1$ as channel of
communication between agents $i$ et $i+m$.  We note $C_{m}$ the
capacity of this channel (it depends only on $m$, the length of the
channel). Following the same derivation as before, we obtain
$$\begin{pmatrix}
\mathbb{P} (X_{i+m}(t+m) = 1) \\ 
\mathbb{P} (X_{i+m}(t+m) = 0)
\end{pmatrix}  = A^{m} \begin{pmatrix}
\beta \\ 
1- \beta
\end{pmatrix}$$
Since $A$ is a symmetric matrix it can be cast in a diagonal form with an
orthonormal basis. The eigenvalues are $\lambda_1=1$ and $\lambda_2 =
1- 2 \epsilon$. Thus, $A^m$ can be expressed as
$$A^{m} = P^T \begin{pmatrix}
1 & 0 \\ 
0 & (1- 2\epsilon)^{m}
\end{pmatrix} P$$
with $P = \dfrac{1}{\sqrt{2}} \begin{pmatrix}
1 & 1 \\ 
1 & -1
\end{pmatrix}$
Thus 
$$A^{m}  = \dfrac{1}{2} \begin{pmatrix}
1+(1- 2 \epsilon)^m & 1-(1- 2 \epsilon)^m\\ 
1-(1- 2 \epsilon)^m & 1+(1- 2 \epsilon)^m
\end{pmatrix} $$
and we obtain a symmetric binary channel of length $m$ with a
probabily of error
$$\epsilon_m = \dfrac{1}{2} (1 - (1-2 \epsilon)^m)$$
Therefore, the capacity of this channel  is
$$C_{m} = 1 + \epsilon_m \log_2 (\epsilon_m) + (1- \epsilon_m) \log_2(1- \epsilon_m)$$

\begin{figure}[htb]
 \psfig{figure=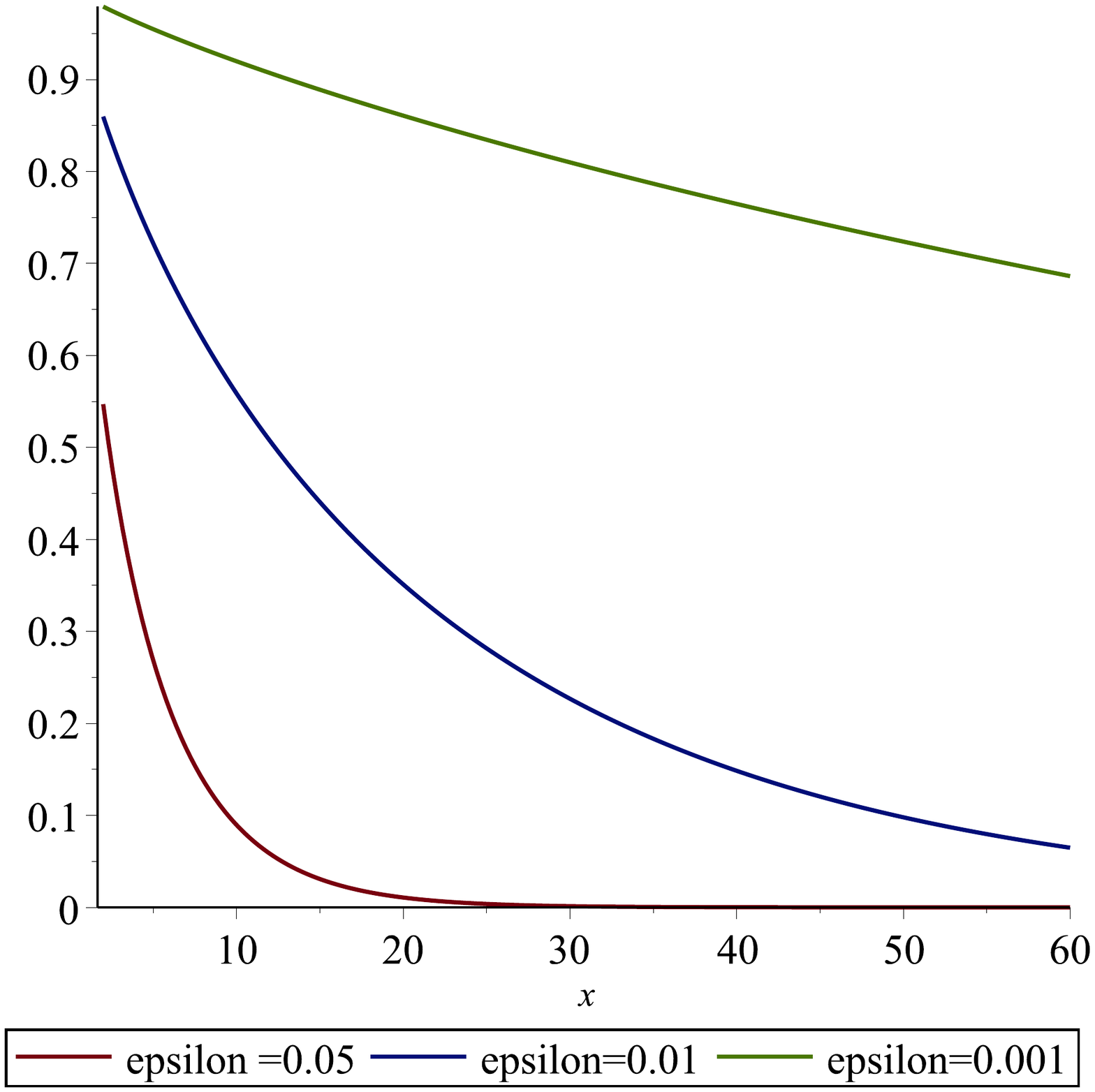,width=.5\textwidth}
 \caption{Capacity $C_m$ of the channel between agent $i$ and $i+m$, as a function of $m$, for different noise levels $\epsilon$. }
\label{fig:capacite}
\end{figure}

We know that the capacity is an upper bound of the mutual information
for each value of $\epsilon$. In Fig.~\ref{fig:capacite}, the
capacity $C_m$ is shown as a function of its length $m$, for different
values of the noise. The fact that the capacity $C_m$ decreases
with $m$ and with the noise, gives another confirmation of the
increasing difficulty to control agent $i+m$ by forcing the vote of
agent $i$. 

\section{Linear voter model and control theory}\label{sect:controle}

The linear voter model analysis here above may be interpreted in terms
of reachability or observability using classical tools from system
(control) theory (see \cite{Antoulas05}, chapter 4 for
an introduction to the control notions used hereafter). Let us
consider again a linear topology with $n+1$ voting agents. In
section~\ref{sect:1D}, we mostly considered the case where agent was
forced to vote 1.  Here we consider a more general case. For $l,m\in \{0, \ldots, n\}$, forcing the vote of  agent $l$
may be considered as a control action, while observing the vote of agent
$m$  may be considered as an output measurement. Since we are
interested in the deviation from 1/2 of the probability to  vote 1
(thus measuring the influence of a forcing action, for
instance), we define these deviations as state space variables
\begin{equation}
\label{eq:statespacevariablesdef}
\tilde{p}_i(t):=p_i(t)-\frac{1}{2}\in\left[ -\frac{1}{2} ; \frac{1}{2} \right]
\end{equation}
for all $t\geq 0$ and $i\in  \{0, \ldots, n\}$. We will consider in the sequel, with no loss of generality, a forcing of agent $0$ vote and an observation of agent $n$ vote, since the influence in the considered linear voter is unidirectional (from left to right). Therefore, the input variable, $\tilde{u}$, and output variable, $\tilde{y}$, will be defined as 
\begin{equation}
\label{eq:IOvariablesdef}
\tilde{u}(t):=p_0(t)-\frac{1}{2} \; ; \; \tilde{y}(t):=p_n(t)-\frac{1}{2}
\end{equation}
Using these state space, input and output variables, the dynamical voter model \eqref{eq:matrix-form} transforms into the state space system
\begin{eqnarray}
\label{eq:linearvoterSSS}
\tilde{\mathbf{p}}(t+1)&=&A\tilde{\mathbf{p}}(t) + \mathbf{b} \tilde{u}(t) \\
\tilde{y}(t)&=& \mathbf{c}^T \tilde{\mathbf{p}}(t) \nonumber
\end{eqnarray}
with the state vector $\tilde{\mathbf{p}}(t):=\left[ \tilde{p}_1(t), \ldots , \tilde{p}_n(t) \right]^T\in \mathbb{R}^n$ and  the internal dynamics matrix (generator) $A$ defined as
\begin{equation}
\label{eq:Adef}
 A := \left( \frac{1}{2} - \epsilon \right) \begin{pmatrix}
1 & 0 & \ldots & \ldots  & 0 \\ 
1 & 1 & 0 & \ldots & 0 \\ 
0 & 1 &1 & \ddots & 0 \\ 
\vdots & \ddots & \ddots & \ddots &0\\ 
0 & \ldots & 0 & 1 & 1
\end{pmatrix}
\end{equation}
The control column matrix $\mathbf{b}$ and observation row matrix $\mathbf{c}^T$ are defined respectively as
\begin{equation}
\label{eq:bcdef}
\mathbf{b}:= \left( \frac{1}{2}-\epsilon \right)
\begin{pmatrix}
1 \\ 
0 \\ 
\vdots \\ 
0
\end{pmatrix} \; \mbox{ and } \;
\mathbf{c}^T:=\begin{pmatrix}
0 & \ldots & 0 & 1
\end{pmatrix}
\end{equation}
For any time $t\geq 0$, any initial probability distribution $\tilde{\mathbf{p}}(0):=\tilde{\mathbf{p}}_0\in \left[-\frac{1}{2} ; \frac{1}{2}\right]^n$ and any control (forcing) signal values $\tilde{u}(t)\in \left[-\frac{1}{2} ; \frac{1}{2}\right]$, the solution $\mathbf{\phi} \left( \tilde{u};\tilde{\mathbf{p}}_0;t\right)$ of the state space equations \eqref{eq:linearvoterSSS} may be written
\begin{equation}
\label{eq:SSSsolution}
\mathbf{\phi} \left( \tilde{u};\tilde{\mathbf{p}}_0;t\right)=A^t\tilde{\mathbf{p}}_0+\sum_{j=0}^{t-1}A^{(t-1)-j}\mathbf{b}\tilde{u}(j)
\end{equation}
Note that the matrix $A$ has a unique eigenvalue $\lambda(A)=(\frac{1}{2}-\epsilon)$, with multiplicity $n$ and such that $\left| \lambda(A) \right| < 1$ (since the noise $\epsilon$ satisfies $0\leq \epsilon <\frac{1}{2}$). Therefore the trajectory \eqref{eq:SSSsolution} is bounded when $t\rightarrow \infty$ and the dynamical system \eqref{eq:linearvoterSSS} is said {\em stable}.

A state $\tilde{\mathbf{p}} \in  \left[-\frac{1}{2} ; \frac{1}{2}\right]^n$ is said {\em unobservable} if the corresponding output can not be distinguished from the output associated with the zero state, that is if 
\begin{equation}
y(t)=\mathbf{c}^T \phi \left( 0;\tilde{\mathbf{p}};t\right) =0
\end{equation}
for all $t\geq 0$ (in the observability analysis, only the free response dynamics is analyzed and $\tilde{u}$ is set to zero). The whole state space system \eqref{eq:linearvoterSSS} is said {\em observable} if the set of unobservable states reduces to $\{ 0 \}$. With the solution \eqref{eq:SSSsolution} and Cayley theorem, it is easy to prove that this is the case if and only if the {\em observability matrix}
\begin{equation}
\label{eq:observabilitymatrixdef}
\mathcal{O}_n=\left[ \mathbf{c} \; ; \; A^T\mathbf{c} \; ; \; \ldots \; ; \; (A^{n-1})^T\mathbf{c} \right]^T
\end{equation}
is full rank or when the {\em infinite observability Gramian}
\begin{equation}
\label{eq:observabilitygramiandef}
W^o:=\lim_{n\rightarrow \infty}{\mathcal{O}_n^T\mathcal{O}_n}=\sum_{k=0}^\infty{(A^k)^T\mathbf{c}\mathbf{c}^TA^k}
\end{equation}
is strictly positive definite. 

The infinite observability Gramian gives additionnal quantitative information about how much the system or a particular state is observable. Indeed, the largest observation energy (i.e. the maximum energy for the output signal) is reached when $t\rightarrow \infty$ and equals
\begin{equation}
\label{eq:outputenergy}
\left\| \tilde{y} \right\|^2_2:=\lim_{t\rightarrow \infty}{\sum_{k=0}^t |\tilde{y}(k)|^2}=\tilde{\mathbf{p}}^TW^o\tilde{\mathbf{p}}
\end{equation}
for any given state space trajectory $\phi \left( 0;\tilde{\mathbf{p}};t\right)$. Therefore, with the appropriate change of state space coordinates, the components of the initial  condition (or subspaces) may be re-ordered, from the less to the most observable ones. If some of the infinite horizon observability Gramian eigenvalue are zero, then the corresponding vector spaces are unobservable. If some of these eigenvalues are small, then initial conditions variations in the corresponding subspaces will cause low energy variations in the output signal. 

In the linear voter model example, rather than measuring the influence of forcing permanently the agent 0 to vote 1 (with a constant input signal $\tilde{u}(t)=\frac{1}{2}, \, \forall t\geq 0$) on the vote of agent $n$, we could instead analyze to effect of considering the initial probability distribution
\begin{equation}
\label{eq:initialprofile}
\tilde{\mathbf{p}}:=\left[ 1, 0, \ldots , 0 \right]^T\in \mathbb{R}^{n+1}
\end{equation}
on agent $n+1$, by measuring the corresponding observation energy. We will consider a long range time horizon $k > n$ for which the influence of the initial state of agent $1$ has reached agent $n+1$ in the line. The last row of matrix $A^k$ may be written (see \ref{Appendix_Jordan}):
\begin{equation}
\label{eq:Akderniereligne}
A^k_{(n+1,\cdot)}:=
\begin{cases}
(\frac{1}{2}-\epsilon)^k\left[ \binom{k}{n} \; \binom{k}{n-1} \; \ldots \; \binom{k}{2} \;k \;1 \right] & \mbox{ when } k\geq n \\
\\
(\frac{1}{2}-\epsilon)^k\left[ 0 \; \ldots \; 0 \; \binom{k}{k} \; \ldots \; \binom{k}{2} \;k \;1 \right] & \mbox{ when } k< n
\end{cases}
\end{equation}
According to definition~\eqref{eq:observabilitygramiandef}, since we
are measuring the vote of agent $n+1$, we get for the components of
the infinite observability Gramian
\begin{equation}
\label{eq:Woelements}
W^o_{i,j}:=\sum_{k=0}^\infty{
A^k_{(n+1,i)}A^k_{(n+1,j)}
}
\end{equation}  
for all $i,j\in\{ 1, \ldots , n\}$. Measuring the influence of the initial vote of agent 1, we start with the initial probability distribution \eqref{eq:initialprofile} and get, for the agent $n+1$, the observation energy
\begin{equation}
\label{eq:outputenergylinearvoter}
\left\| \tilde{y} \right\|^2_2:=\sum_{k=0}^\infty{\left( A^k_{(n+1,1)}\right)^2}=\sum_{k=0}^\infty{ (\frac{1}{2}-\epsilon)^{2k}
\left( A^k_{(n+1,1)}\right)^2}
\end{equation}
With equation \eqref{eq:Akderniereligne}, one gets
\begin{eqnarray}
\label{eq:outputenergylinearvoter}
\left\| \tilde{y} \right\|^2_2&=&\sum_{k=n}^\infty{ (\frac{1}{2}-\epsilon)^{2k}
\binom{k}{n}^2} \\ &=& \frac{1}{(n!)^2}(\frac{1}{2}-\epsilon)^{2n} \sum_{p=0}^\infty{ (\frac{1}{2}-\epsilon)^{2p}
\left( \frac{(p+n)!}{p!} \right)^2} \nonumber
\end{eqnarray}
Using the lower bound
\begin{equation}
(p+1)^n < \frac{(p+n)!}{p!}
\end{equation}
one gets
\begin{equation}
\label{lowerbound}
\frac{1}{((n-1)!)^2}(\frac{1}{2}-\epsilon)^{2n} \frac{4(3-2\epsilon)}{(1+2\epsilon)^3} < \left\| \tilde{y} \right\|^2_2
\end{equation}
On the other hand, since
\begin{eqnarray}
& & \sum_{p=0}^\infty{ (\frac{1}{2}-\epsilon)^{2p}
\left( \frac{(p+n)!}{p!} \right)^2} \nonumber \\
& & \leq  \left( \sum_{p=0}^\infty{ (\frac{1}{2}-\epsilon)^{p}
\frac{(p+n)!}{p!}} \right)^2  \nonumber \\
&= & \left( \sum_{k\geq n}{ (\frac{1}{2}-\epsilon)^{k-n}
 k(k-1)\ldots (k-(n-1))} \right)^2  \nonumber \\
&=&\left( \frac{n!2^{n+1}}{(1+2\epsilon)^{n+1}} \right)^2
\end{eqnarray}
we get the following upper bound for the observation energy
\begin{equation}
\label{upperbound}
\left\| \tilde{y} \right\|^2_2 < \frac{4(1-2\epsilon)^{2n}}{(1+2\epsilon)^{2n+2}}
\end{equation}
It is worthwile to notice how this upper bound behaves with the number
of agents along the line and with the noise $\epsilon$. For instance,
the upper bound~\eqref{upperbound} decreases with the number of agents
and the corresponding observation energy is divided by two when $k$
supplementary agents are added in the line, with
\begin{equation}
k \geq \frac{1}{2} \frac{1}{\log_2{\left( \frac{1+2\epsilon}{1-2\epsilon} \right)}} = \frac{\ell_c}{2}
\end{equation}
When the noise increases, the observation energy upper bound decreases
 \begin{equation}
\left\| \tilde{y} \right\|^2_2 = \mathcal{O} \left( (1 - 2\epsilon)^{2n} \right) \rightarrow 0 \mbox{ as } \epsilon \rightarrow \left( \frac{1}{2} \right)^{-}
\end{equation}
The lower bound \eqref{lowerbound} decreases similarly, with the same order, when the noise decreases. However it decreases much faster with the number of agents in the voter line since this lower bound for the observation energy is divided by $\left( \frac{1-2\epsilon}{2n} \right)^2$ when only one agent is added to the $n$ previous ones.

Note that we performed the observability analysis on the linear voter model. We could as well develop the dual reachability analysis for the same example. In this analysis, the initial condition is assumed to be zero and one analyzes the forced solution of the state space model \eqref{eq:linearvoterSSS}. More specifically, one could be interested in its reachability property. A state $\tilde{\mathbf{p}}$ is said {\em reachable} when there is a an input signal $\tilde{u}(t)$ such that 
\[
\lim_{t\rightarrow \infty}{\phi \left( \tilde{u}, \mathbf{0}, t \right)}=\tilde{\mathbf{p}}
\]
It may be proved (see, e.g. \cite{Antoulas05}) that, among those input signals which can reach the state $\tilde{\mathbf{p}}$ from a zero intial condition,  the one with minimum energy may be written as
\[
\left\| \tilde{u}^{*} \right\|_2^2=\tilde{\mathbf{p}}^T(W^c)^{-1}\tilde{\mathbf{p}}
\]
where the {\em infinite reachability Gramian} $W^c$ is defined as
\begin{equation}
\label{eq:reachabilitygramiandef}
W^c:=\sum_{k=0}^\infty{A^k\mathbf{b}\mathbf{b}^T(A^T)^k}
\end{equation}
Therefore, a reachability Gramian analysis may be used to compute the forcing of agent $1$ with minimal energy requested to  reach a state $\tilde{\mathbf{p}}$ where all agents in the line vote 1, that is such that $\tilde{{p}}_i=1$, for all $i\in{1,\ldots, n+1}$. However, in this case, it would be necessary to compute the sum of all the elements in $(W^c)^{-1}$, which is a much more involved computation than the one performed for the observability analysis. Besides, the duality between reachability and observability for linear systems \cite{Antoulas05} and the particular topology of the linear voter model lead us to the conjecture that the reachability analysis would not bring any new result fundamentally different from the ones obtained through the observability analysis.

\section{Conclusions}\label{sect:conclusion}

In this paper we show that time delayed mutual- and multi-informations are
promising tools to better grasp the behavior of a dynamical system
on complex networks. In particular it can be used to determine the
most influential degrees of freedom and the most observable variables.
This knowledge can be obtained without perturbing the system, by just
probing its behavior.

We claim that influential nodes are
those that are the most interesting to control or monitor to (i) force
a system to reach a given target, or (ii) to have a proxy giving an
information on the state of the entire system.

We illustrated our approach in a simple stochastic dynamical model on
a graph, a so-called voter model, where agents iteratively adapt their opinion
to that of the majority of their neighbors, with however a given noise
level. We first discussed the case of a general scale-free
topology, where only numerical results can be obtained. Then we
consider a 1D topology for which analytical results can be
obtained. There, we rigorously showed that the influence of an agent
on the entire system can be equivalently measured by actually forcing
its behavior, or, in a non-intrusive way, by measuring
the time delayed multi-information of this agent with respect to the
rest of the system. In particular, we proposed the concept of a
control length, which indicates a characteristic distance above which
the influence of a controlled agent fades exponentially. 

The link with classical control theory has been proposed and the
control length has been related to the reachability Gramian, thus
indicating that the cost of control becomes intractable at large
distance.  The importance of the noise is clearly shown as being a
central element in the possibility of observing or controlling a
system, as opposed to previous literature that claimed that a
causality path was sufficient to achieve control~\cite{Liu11}.

As an additional link of our approach to existing concepts, we showed
that controllability can also be considered in the framework of the
capacity of communication channel, as defined in information theory by
Shannon. We showed that this capacity drops as agent are separated by
a distance above the control length.

In a forthcoming paper we will apply our approach to other complex
systems, in particular those for which the underlying dynamics and
topology of interaction are not known. We already obtained (not shown
here) that the time delayed multi-information can be used to infer the
topology of the graph of Fig.~\ref{fig:graph_colored}. Further, we
want to use the concept of observability as a way to detect early
warning signal of possible tipping points in a complex dynamical
system. In simple words, we want to analyze the idea that the most
influential degree of freedom is the best variable to observe to know
in advance if a given system is likely to move to another regime.
These nodes being the most influential ones, we can argue that their
evolution will dictate the evolution of the other variables.

\section*{Acknowledgment}

We thank Gregor Chliamovitch and Alex Dupuis for initiating several of
the ideas developed in this paper, during the FP7 project SOPHOCLES
(2012-2015), and for the reading of the manuscript.
 
\bibliographystyle{plain}

\bibliography{References_info_control}

\begin{thebibliography}{1}

\bibitem{Antoulas05}
Athanasios~C Antoulas.
\newblock {\em Approximation of large-scale dynamical systems}, volume~6.
\newblock Siam, 2005.

\bibitem{Barabasi2000}
Albert-L{\'a}szl{\'o} Barab{\'a}si, R{\'e}ka Albert, and Hawoong Jeong.
\newblock Scale-free characteristics of random networks: the topology of the
  world-wide web.
\newblock {\em Physica A: statistical mechanics and its applications},
  281(1-4):69--77, 2000.

\bibitem{Bollobas2003}
B{\'e}la Bollob{\'a}s and Oliver~M Riordan.
\newblock Mathematical results on scale-free random graphs.
\newblock {\em Handbook of graphs and networks: from the genome to the
  internet}, pages 1--34, 2003.

\bibitem{RevModPhys.81.591}
Claudio Castellano, Santo Fortunato, and Vittorio Loreto.
\newblock Statistical physics of social dynamics.
\newblock {\em Rev. Mod. Phys.}, 81:591--646, May 2009.

\bibitem{cover2012}
Thomas~M Cover and Joy~A Thomas.
\newblock {\em Elements of information theory}.
\newblock John Wiley \& Sons, 2012.

\bibitem{BC-galam:98}
S.~Galam, B.~Chopard, A.~Masselot, and M.~Droz.
\newblock Competing species dynamics: Qualitative advantage versus geography.
\newblock {\em Eur. Phys. J. B}, 4:529--531, 1998.

\bibitem{Liu11}
Y.-Y. Liu, J.-J. Slotine, and A.-L. Barabasi.
\newblock Controllability of complex networks.
\newblock {\em Nature}, 473:167, 2011.

\bibitem{pearl09}
J.~Pearl.
\newblock {\em Causality: Models, Reasoning and Inference}.
\newblock Cambridge University Press, 2009.

\end{thebibliography}

\onecolumn

\appendix \label{Appendix}

\section{Explicit evaluation of $A^p$}\label{Appendix_Jordan}


In  \ref{eq_Jordan}, we need to calculate $A^p$ with 
$$A= ( \dfrac{1}{2} - \epsilon ) \begin{pmatrix}
1 & 0 & \ldots & \ldots  & 0 \\ 
1 & 1 & 0 & \ddots & \vdots \\ 
0 & 1 &1 & \ddots & 0 \\ 
\vdots &  & \ddots & \ddots &0\\ 
0 & \ldots & 0 & 1 & 1
\end{pmatrix}$$
We have $A=I_n+C$ with 
$$C = \begin{pmatrix}
0 & 0 & \ldots & \ldots  & 0 \\ 
1 & 0 & 0 & \ddots & \vdots \\ 
0 & 1 &0 & \ddots & 0 \\ 
\vdots &  & \ddots & \ddots &0\\ 
0 & \ldots & 0 & 1 & 0
\end{pmatrix}$$
$C$ is a nilpotent matrix and $\forall k \in \mathbb{N},\ A^k = \sum_{p=0}^k  \binom{k}{p} C^p$. Therefore, for $p \geqslant n-1$:
$$A^p= ( \dfrac{1}{2} - \epsilon )^p \begin{pmatrix}
    1 & 0 & \ldots & \ldots  & 0 \\ 
    \binom{p}{1}   & \ddots & \ddots &  & \vdots \\ 
     \binom{p}{2} & \ddots & \ddots & \ddots & \vdots \\ 
     \vdots & \ddots& \ddots & \ddots & 0 \\ 
     \binom{p}{n-1} & \ldots & \binom{p}{2} & \binom{p}{1} & 1
     \end{pmatrix} 
      $$
and for $ p< n-1$: 
$$A^p = ( \dfrac{1}{2} - \epsilon )^p \begin{pmatrix}
     1 & 0 & \ldots & \ldots & \ldots & \ldots & \ldots & 0 \\ 
      \binom{p}{1} & 1 & \ddots &  &  &  &  & \vdots \\ 
      \binom{p}{2} & \binom{p}{1} & \ddots & \ddots &  &  &  & \vdots \\ 
      \vdots & \binom{p}{2} & \ddots & \ddots & \ddots &  &  & \vdots \\ 
      \binom{p}{p} &  & \ddots &  & \ddots & \ddots &  & \vdots \\ 
      0 & \ddots &  & \ddots & \ddots & \ddots & 0 & \vdots \\ 
      \vdots & \ddots & \ddots &  & \ddots & \ddots & \ddots & 0 \\ 
      0 & \ldots & 0 & \binom{p}{p} & \ldots & \binom{p}{2} & \binom{p}{1} & 1
      \end{pmatrix} $$ 

\section{Accuracy and confidence for the numerical evaluation of probability distributions} \label{appendix:Central_limit}

Let us consider an attribute of the members of a population which
appears with probability $p$.  For a sample of size $n$ drawn in this
population, let $F_n$ be the random variable equal to the proportion
of those elements  having this attribute. According to the
Moivre-Laplace theorem, the quantity $\frac{F_n-p}{\sqrt{p(1-p)/n}}$
converges in distribution to a Gaussian distribution
\begin{align*}
\mathbb{P} (  \mid F_n-p \mid \leqslant \epsilon ) = 1- \alpha & \Leftrightarrow  \mathbb{P} \Big( \frac{ \mid F_n-p \mid }{\sqrt{\frac{p(1-p)}{n}}} \leqslant \frac{\epsilon}{\sqrt{\frac{p(1-p)}{n}}} \Big) = 1- \alpha    \\ 
& \Leftrightarrow   \frac{\epsilon \sqrt{n}}{\sqrt{p(1-p)}} = t_{1- \alpha/2} \\
& \Leftrightarrow \epsilon = t_{1- \alpha/2} \sqrt{\frac{p(1-p)}{n}} \\
\end{align*}
where $t_{1- \alpha/2}$ is the real number defined by $\mathbb{P}(X \leqslant t_{1- \alpha/2}) = \alpha$ with $X \sim \mathcal{N}(0,1)$. As $p \in [0,1]$ and $p(1-p) \leqslant 0.25$, therefore  $\epsilon \leqslant t_{1- \alpha/2} \frac{1}{2 \sqrt{n}}$. For $N=10^5$ and $\alpha =0,05$, we have $t_{1- \alpha /2} = 1.96$, we obtain  an approximation value of $p$ with a precision of $0.03$,  with a  risk of $5 \%$.

\section{Algorithms for the computations of mutual and multi-information}\label{Algo1}

\subsection{Mutual information}
We consider a scale free graph  $G$, with $n$ agents. To compute the $\tau$-delayed mutual information between 2 agents $i$ and $j$, we generate $N$ runs.
For every run, we have a matrix $S$ defined by: for $i \in \llbracket 1,n \rrbracket$, and for $j \in \llbracket 1,t+ \tau \rrbracket$, such that $S[i][j]$ is the state of the agent $i$ at the moment $j$.

We use $4$ $n\times n$ matrix (N00, N01, N10 and N11), initialized to zeros. For every run, we compare the vote
of the agent $i$ at the time $t$ and the vote of the agent $j$ at the
time $t+\tau$\\
\vskip .2cm\noindent
for $i$ from $1$ to $n$\\
for $j$ from $1$ to $n$\\
if $S[i][t]=0$ and $S[j][t+\tau]=0$ then $N00[i][j]++$ endif\\
if $S[i][t]=0$ and $S[j][t+\tau]=1$ then $N01[i][j]++$ endif\\
if $S[i][t]=1$ and $S[j][t+\tau]=0$ then $N10[i][j]++$ endif\\
if $S[i][t]=1$ and $S[j][t+\tau]=1$ then $N00[i][j]++$ endif\\
end for\\
end for\\
\vskip .2cm\noindent We then compute $w_{i,j}(t,\tau)$, the
$\tau$-delayed mutual information at time $t$ between agents $i$ and
$j$, according to  definition~(\ref{info_mutual}). We obtain
\begin{eqnarray}
  \forall (i,j),\ w_{i,j} (t,\tau) &=& \dfrac{N00[i][j]}{N} \log_2 \Big(  \dfrac{N00[i][j] \times N}{(N00[i][j]
    +N10[i][j])\times (N00[i][j]+N10[i][j])} \Big)
 \nonumber\\
&&+
\dfrac{N01[i][j]}{N} \log_2 \Big(  \dfrac{N01[i][j] \times N}{(N00[i][j]+N01[i][j])\times (N01[i][j]+N11[i][j])} \Big)
 \nonumber\\
&&+
\dfrac{N00[i][j]}{N} \log_2 \Big(  \dfrac{N10[i][j] \times N}{(N10[i][j]+N[i][j])\times (N00[i][j]+N10[i][j])} \Big)
 \nonumber\\
&&+
 \dfrac{N11[i][j]}{N} \log_2 \Big(  \dfrac{N11[i][j] \times N}{(N10[i][j]+N11[i][j])\times (N01[i][j]+N11[i][j])} \Big)
 \nonumber\\
\end{eqnarray}

\subsection{Multi-information}
To compute the delayed multi-information, as for the delayed mutual
information, we execute $N$ runs, and for every run, we compute the
state matrix $S$.\\ We use two $n\times n$ matrix, $N0$
and $N1$ ($n$ is the number of agents) defined by : $\forall
(i,j),\ N0[i][j]$ is equal to the number of runs where the vote
of the agent $i$ is $0$ and the number of agents who voted $1$ is
$j-1$ at time $t+\tau$.\\ $\forall (i,j),\ N1[i][j]$ is
equal to the number of runs where the vote of the agent $i$ is $1$ and
the number of agents (without the agent $i$) who voted $1$ is
$j-1$.\\ These matrices give us the probability distribution of the
couple of random variable $(X_i(t),Y_i(t+\tau))$ that we have in the
definition of the delayed multi-information, eq.~(\ref{info_multi}).

\end{document}